\documentclass[pdftex,twocolumn,epjc3]{svjour3}          

\RequirePackage[T1]{fontenc}

\smartqed  

\usepackage{graphicx}
\usepackage{mathptmx}      
\usepackage{flushend}
\usepackage{amsmath}
\usepackage[normalem]{ulem}
\usepackage{subcaption,arydshln}
\usepackage[numbers,sort&compress]{natbib}
\usepackage[colorlinks,citecolor=blue,urlcolor=blue,linkcolor=blue]{hyperref}

\journalname{Eur. Phys. J. C}

\begin{document}

\title{CNN on `Top'}

\subtitle{In Search of Scalable \& Lightweight Image-based Jet Taggers}

\author{Rajneil Baruah\thanksref{e1,addr1}
        \and
        Subhadeep Mondal\thanksref{e2,addr1} 
        \and 
        Sunando Kumar Patra\thanksref{e3,addr2}
        \and
        Satyajit Roy\thanksref{e4,addr2}
}

\thankstext{e1}{e-mail: rajneilb.physics@gmail.com }
\thankstext{e2}{e-mail: subhadeep.mondal@bennett.edu.in}
\thankstext{e3}{e-mail: sunando.patra@gmail.com}
\thankstext{e4}{e-mail: roy.satya05@gmail.com}

\institute{Department of Physics, SEAS, Bennett University, Greater Noida, Uttar Pradesh, 201310, India\label{addr1}
          \and
          Department of Physics, Bangabasi Evening College, Kolkata, 700009, West
Bengal, India\label{addr2}
          }

\date{Accepted: 10 October 2026}

\maketitle

\begin{abstract}
While Transformer-based and standard Graph Neural Networks (GNNs) have proven to be the best performers in classifying different types of jets, they require substantial computational power. We explore the scope of using a lightweight and scalable version of \textit{EfficientNet} architecture,  along with global features of the jet. The end product is computationally inexpensive but is capable of competitive performance. We showcase the efficacy of our network in tagging top-quark jets in a sea of other light quark and gluon jets. The work also sheds light on the importance of global features for both the accuracy and the apparent redundancy of the network's complexity.
\end{abstract}

\section{Introduction}
\label{sec:intro}

As we enter the high-energy (HE) and high luminosity (HL) phase of the Large Hadron Collider (LHC), the study of jet physics has become more relevant than ever. `Fat jet tagging' is a critical task for identifying the hadronic decays of massive particles, such as the Higgs boson, top quark, or a $W$ or $Z$-boson that are produced with high momentum or `boost' in a particular direction. As a result of this boost, the decay products of these standard model (SM) particles are highly collimated and can be contained within a single large-radius jet, which we call a `fat jet'. Scenarios like this can easily occur in various beyond the standard model (BSM) frameworks with vector-like quarks \cite{Yang:2025ktj,Ghosh:2025gdq,Banerjee:2024zvg,Choudhury:2021nib}, leptoquarks \cite{Sahoo:2025kdj,Bhaskar:2024snl,Ghosh:2025gue,Chandak:2019iwj,Diaz:2017lit},  heavy right-handed neutrinos \cite{Chakraborty:2018khw,Padhan:2022fak,Abdullahi:2022jlv,Bandyopadhyay:2022mej}, supersymmetric particles with $R$-parity violating couplings \cite{Baruah:2024wrn,Sahu:2025uop,Dreiner:2023bvs,Dreiner:2025kfd}, to name a few. Among the above-mentioned SM particles, top quark holds a unique status. It happens to be the heaviest fundamental particle we know thus far, and unlike other quarks, it decays before hadronization. The heavy mass of the top quark means its coupling strength with the Higgs boson is the strongest. Hence, precise measurements of top quark properties provide a consistency check of the SM. Even subtle deviations from the SM predictions for top quark interactions could hint at new physics. Moreover, the top quark mass plays a crucial role in determining the stability of the electroweak vacuum. Though the current measurements suggest that our vacuum is `metastable' \cite{Hiller:2024zjp,Degrassi:2012ry, Buttazzo:2013uya}, this conclusion is highly sensitive to the precise values of the top quark measurements \cite{Alekhin:2012py,Elias-Miro:2011sqh, Espinosa:2015kwx}.
Given that even subtle deviations in these properties could hint at new physics or alter our understanding of vacuum stability, the ability to precisely isolate top quark events is important. Consequently, identifying top quark jets as \textit{fat jets}—distinguishing them from the overwhelming background of light quark and gluon jets—emerges as a critical task.

Machine learning (ML) has revolutionized this field \cite{Sahu:2024fzi, Bhattacherjee:2022gjq, Mikuni:2025tar} by moving beyond traditional methods that relied on high-level, physics-inspired features like jet mass, jet radius, $n$-subjettiness, energy correlation functions, etc \cite{Thaler:2010tr, Thaler:2011gf, Stewart:2010tn, Larkoski:2013eya, Moult:2016cvt, Komiske:2017aww, Abdesselam:2010pt, Kogler:2018hem, Altheimer:2012mn, Altheimer:2013yza, Adams:2015hiv, Marzani:2019hun, Larkoski:2017jix}. Deep learning models, on the other hand, instead of using simple, hand-crafted features, train themselves to represent the jets as points in a vector space of features best suited for the subsequent classification. The inputs to these models include the four-momentum components of each constituent particle, global features that carry the kinematical properties of the jet, and particle tracking IDs whenever available. From the ML perspective, the jet physics task at hand can be divided into two broad categories: classification and generation. Both of these `downstream' tasks depend on representing the raw, low-level particle or detector data associated with a jet as points in a dense, continuous vector space (feature space), ready to be effectively processed by any ML algorithm. This process, known as `embedding/encoding,' allows the model to learn a more comprehensive and nuanced representation of the substructure of a jet, which can be utilized in subsequent downstream tasks. 

Different types of neural networks have been developed over the last couple of decades, each tailor-made to embed input tensors of a specific type. While a Convolutional neural network (CNN) uses $2D$ convolutions on an image to do this, a recurrent neural network (RNN) is best suited for data presented as sequences, e.g., audio files, time series, and natural language; a graph neural network (GNN) works best on data with distinct global, node/vertex, and edge features, and so on. These models vary widely in their architectures, with completely different workflows. As one type of data can be represented in multiple ways, these models can be applied independently to them with varying performance measures. For example, pixels of a colored digital image can be represented as vectors, multi-channel $2D$ tensors, graphs, or even fixed-length point clouds. Historically, the calorimeter energy deposits of the constituent particles in a jet were first represented as a $2D$ image/grid, with pixel intensity representing their respective energy/$p_T$, making CNNs—followed by a dense/linear network (colloquially known as multi-layer perceptrons or MLPs)—the default networks for jet-embedding and subsequent classification \cite{Baldi:2016fql, Kagan:2020yrm, Kasieczka:2017nvn, Macaluso:2018tck, Kasieczka:2019dbj, Diefenbacher:2019ezd, Bollweg:2019skg, Butter:2017cot}. 

A more direct approach to embedding is to represent the jet as a point cloud of its constituent particles. Each `point' is a particle with its own set of features, e.g., four momentum components, particle IDs, charges, and tracking information. GNNs \cite{Shlomi:2020gdn, Duarte:2020ngm, Thais:2022iok} are particularly well-suited for this representation. The jet is now treated as a graph where each particle is a `node', and the connections between them are `edges' \cite{Komiske:2018cqr, Qu:2019gqs, Bogatskiy:2020tje, Gong:2022lye, Bogatskiy:2023nnw, Bogatskiy:2022czk, Bogatskiy:2023fug}. 
Constituent information can also be represented as a varying length sequence, especially if it is ordered based on a certain property, e.g., by their transverse momentum $p_T$. RNNs—originally developed for processing text or speech—can then be used. The advent of the self- and cross-attention mechanisms gave us the famously parallelizable and scalable \textit{Transformer} architecture \cite{DBLP:journals/corr/VaswaniSPUJGKP17}. Originally used to process and generate natural language sequences, these models can be thought of as special variations of GNNs \cite{JoshiTrans2GNN}, and can process jets either as sequences of constituents or unordered point clouds.

One jarring issue that plagues the existing networks with the highest accuracy in identifying various jets is the computational cost. Achieving high accuracy boils down to extracting maximal information from the jet representation. It can either be gleaned by formidably large networks or can be preprocessed and fed to a smaller network later. The former is the path taken by transformer-based models \cite{Qu:2022mxj,Mikuni:2021pou,Wu:2024thh} and more costly foundational models like \textit{Omnilearn(ed)} \cite{Mikuni:2025tar,Mikuni:2024qsr,Bhimji:2025isp}. Even these, when striving for higher accuracies, bring separately extracted features to the networks `attention'\footnote{
Not adding pairwise information to the multi-head attention of the (\textit{Par-T}) network results in a $1.2\%$ drop in accuracy and a $20-30\%$ drop in background rejection \cite{Qu:2019gqs,Qu:2022mxj}.}. In the latter approach, when jets are considered as particle clouds or graphs, creating a fully connected graph is prohibitively costly in terms of time and computation, mainly because the pairwise interactions (edge information\footnote{i.e., the abstract message-passing, not physical QCD interactions.}) are numerous for a jet with a large number of constituents. These challenges are either bypassed by keeping only the nearest neighbor interactions (\textit{ParticleNet} \cite{Qu:2019gqs}), or by representing data in a geometric algebra over space-time, equivariant under physics-motivated (Lorentz) transformations \cite{Bogatskiy:2022czk,Bogatskiy:2023nnw,Gong:2022lye}. The newer architectures combine both of these approaches \cite{Spinner:2024hjm,Brehmer:2024yqw}. Even running these state-of-the-art models on single-GPU machines becomes an ordeal, let alone training them from scratch. With the rising awareness of the virtues of Transfer Learning, more models are now pre-trained on larger datasets with multiple classes, which automatically requires significant model complexity. This considerably increases their performance, but the downside is that the cost of retraining increases many times over for specific downstream tasks.


Our objective in this work is to revisit computationally low-cost CNNs, as they implicitly utilize local, pairwise information (albeit spatially, through images) to some extent. Depending on the knowledge that the inclusion of global jet features enhances tagging efficiency \cite{Bhattacherjee:2022gjq}, we explore the possibility of constructing a lightweight yet efficient network by incorporating global features with the images. Although computationally lightweight, smaller CNNs such as LeNet \cite{lenetPaper} cannot achieve competitive accuracies. Extracting more information here is equivalent to increasing image resolution and/or model complexity (depth/width). Although CNNs become notoriously computationally costly when scaled up (e.g., ResNet \cite{ResNet1}), recent developments, such as compound scaling \cite{EffNetV1, EffNetV2}, combined with depthwise and pointwise convolutions\cite{MobileNetV1,MobileNetV2,MobileNetV3,MobileNetV4,MNASnet}, resulted in networks that reach higher accuracies at a fraction of the cost. As all versions of \textit{EfficientNet} were originally designed for images with higher resolution, we have built lightweight versions of those, while following the same scaling rules. We aim to explore their capabilities with and without the externally provided global features, while using LeNet as a benchmark for comparison throughout.


The paper is organized as follows. In Section~\ref{sec:method} we provide a detailed account of our methodology. This includes machine configuration, training data preparation, model architectures, and training procedure. In Section~\ref{sec:result} we present our results. Finally, in Section~\ref{sec:concl}, we summarize our work and draw conclusions. 

\begin{figure*}[t!]
    \centering
    \begin{subfigure}[t]{0.24\textwidth}
        \centering
        \includegraphics[height=4cm]{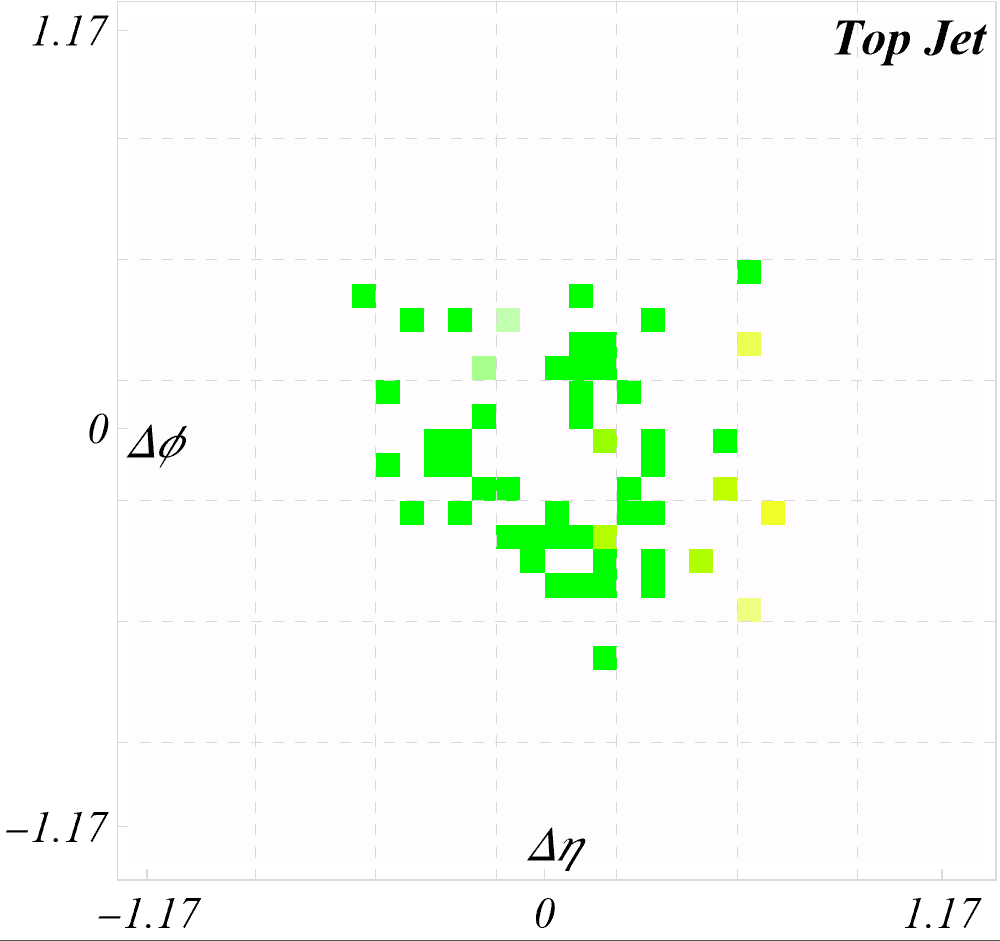}
        \caption{$35\times 35$}
        \label{fig:topjetorig}
    \end{subfigure}~ 
    \begin{subfigure}[t]{0.24\textwidth}
        \centering
        \includegraphics[height=4cm]{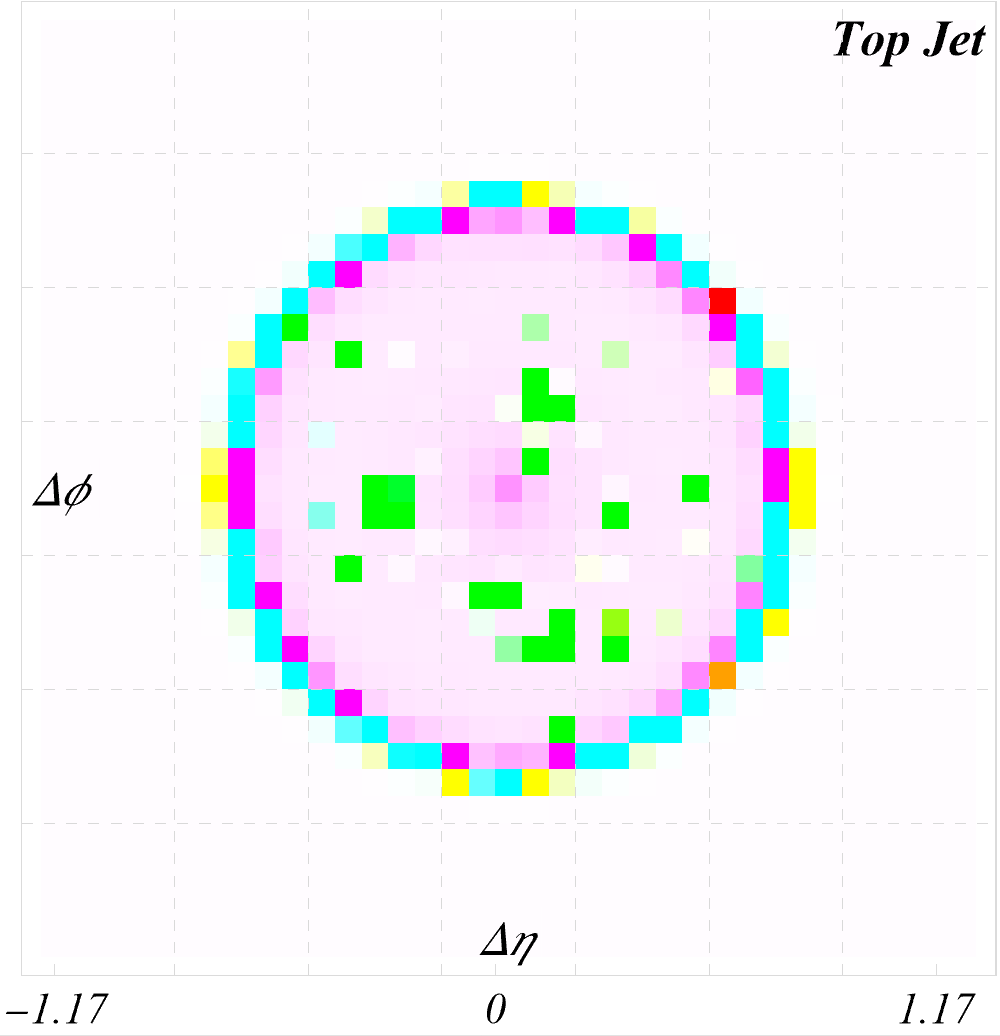}
        \caption{$35\times 35$, Standardized}
        \label{fig:topjetstandard}
    \end{subfigure}~
    \begin{subfigure}[t]{0.24\textwidth}
        \centering
        \includegraphics[height=4cm]{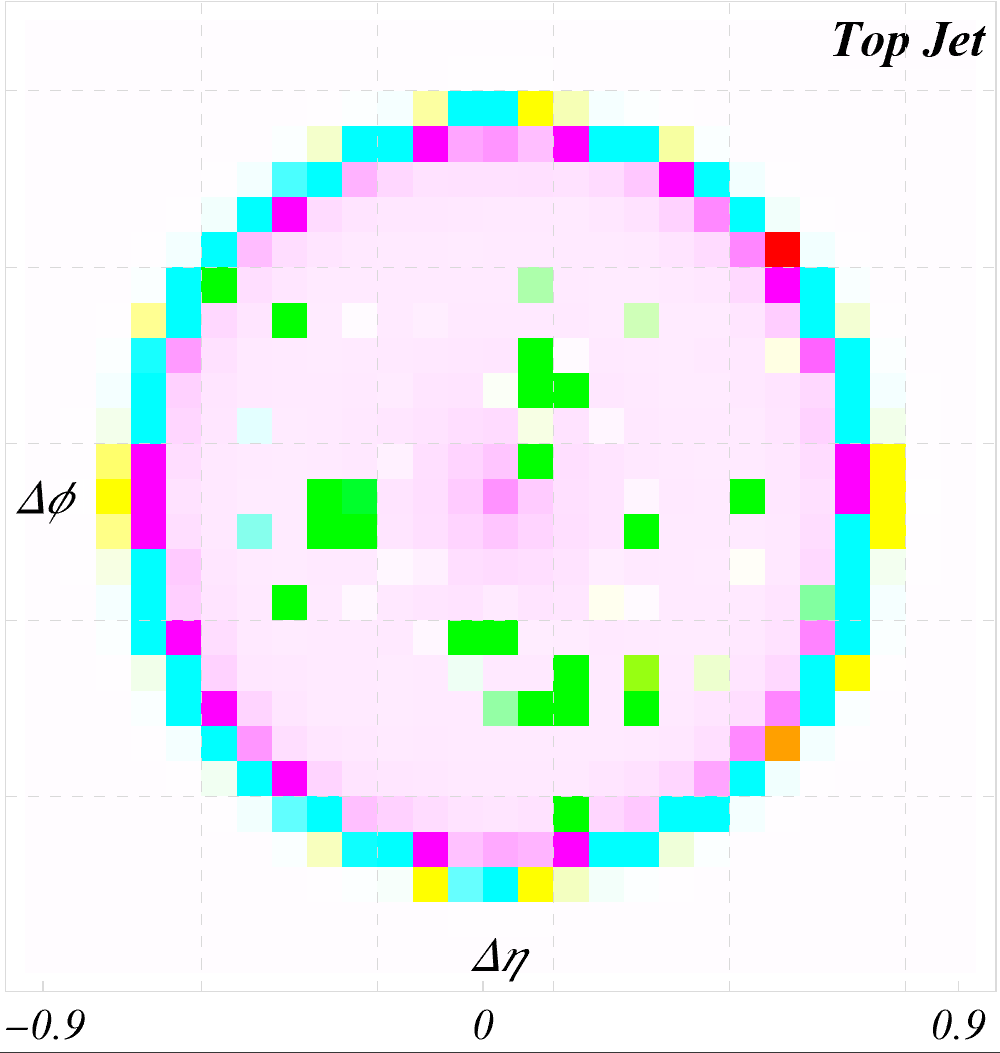}
        \caption{$35 \xrightarrow{Crop} 28$, Standardized}
        \label{fig:topjet32c28}
    \end{subfigure}~ 
    \begin{subfigure}[t]{0.24\textwidth}
        \centering
        \includegraphics[height=4cm]{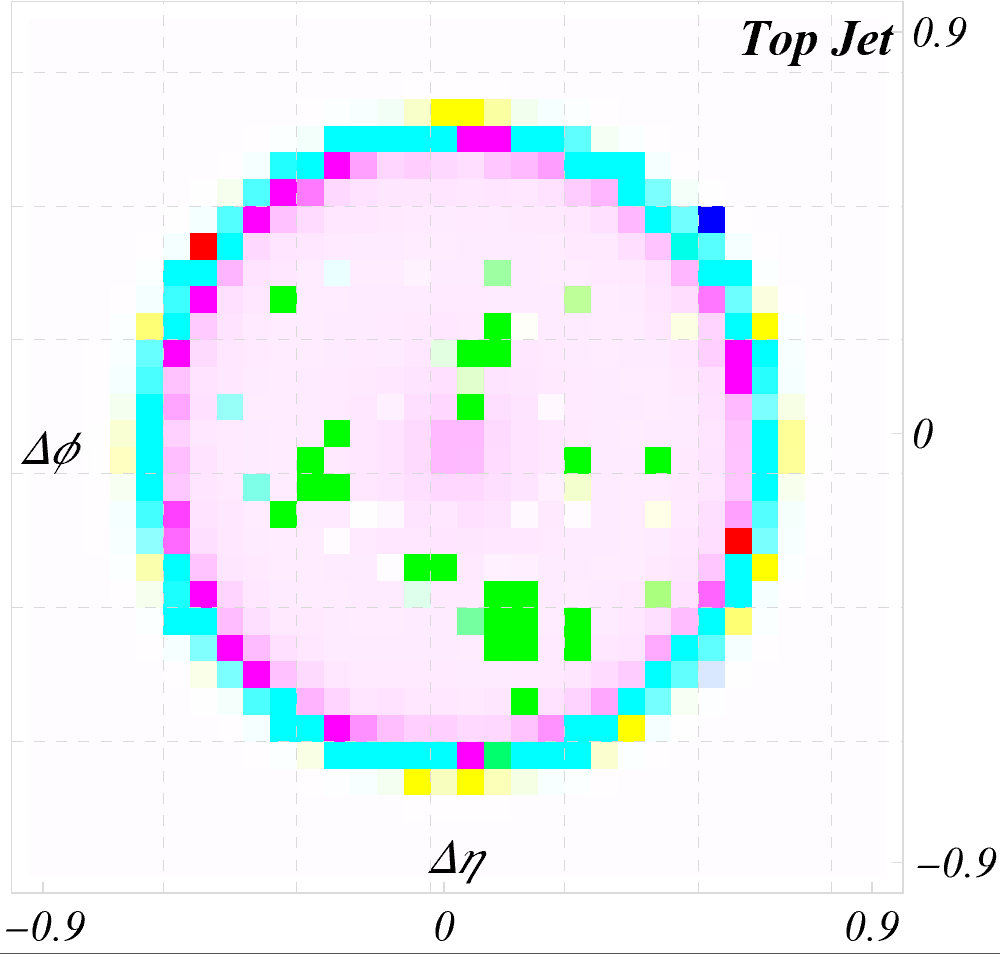}
        \caption{$40 \xrightarrow{Crop} 32$, Standardized}
        \label{fig:topjet40c32}
    \end{subfigure}\\
    \begin{subfigure}[t]{0.24\textwidth}
        \centering
        \includegraphics[height=4cm]{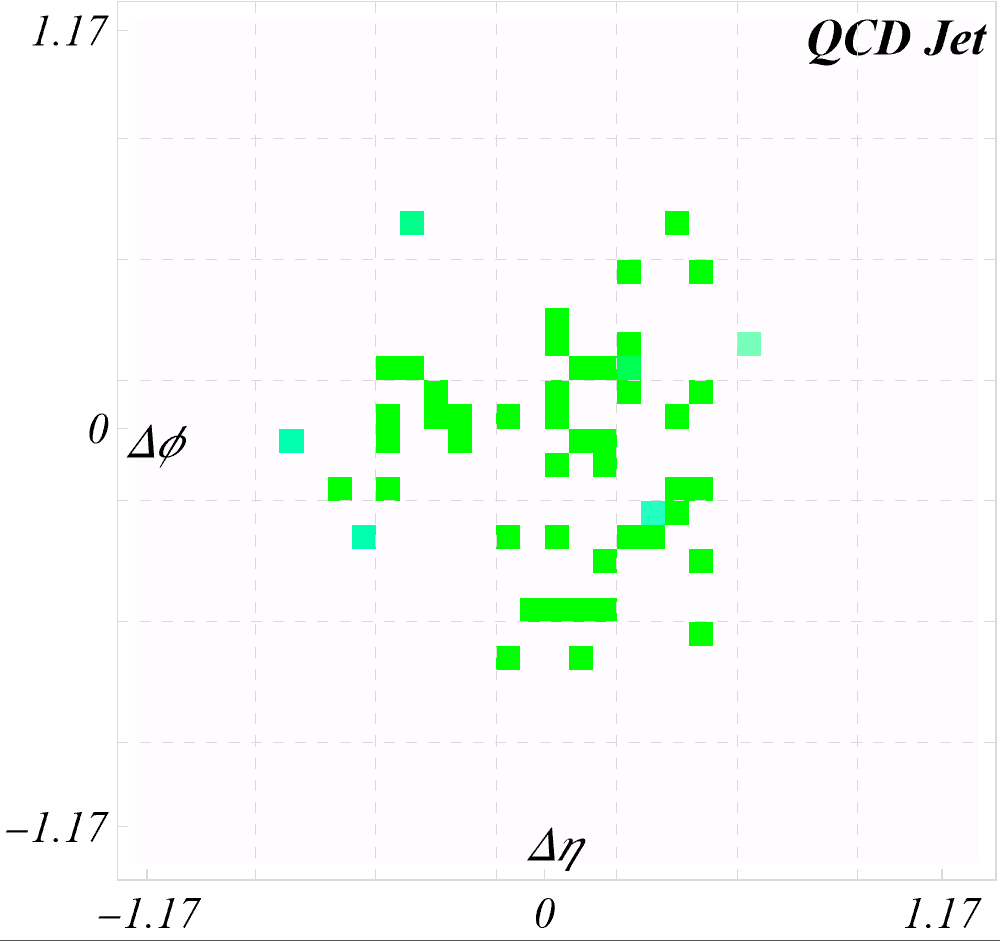}
        \caption{$35\times 35$}
        \label{fig:qcdjetorig}
    \end{subfigure}~ 
    \begin{subfigure}[t]{0.24\textwidth}
        \centering
        \includegraphics[height=4cm]{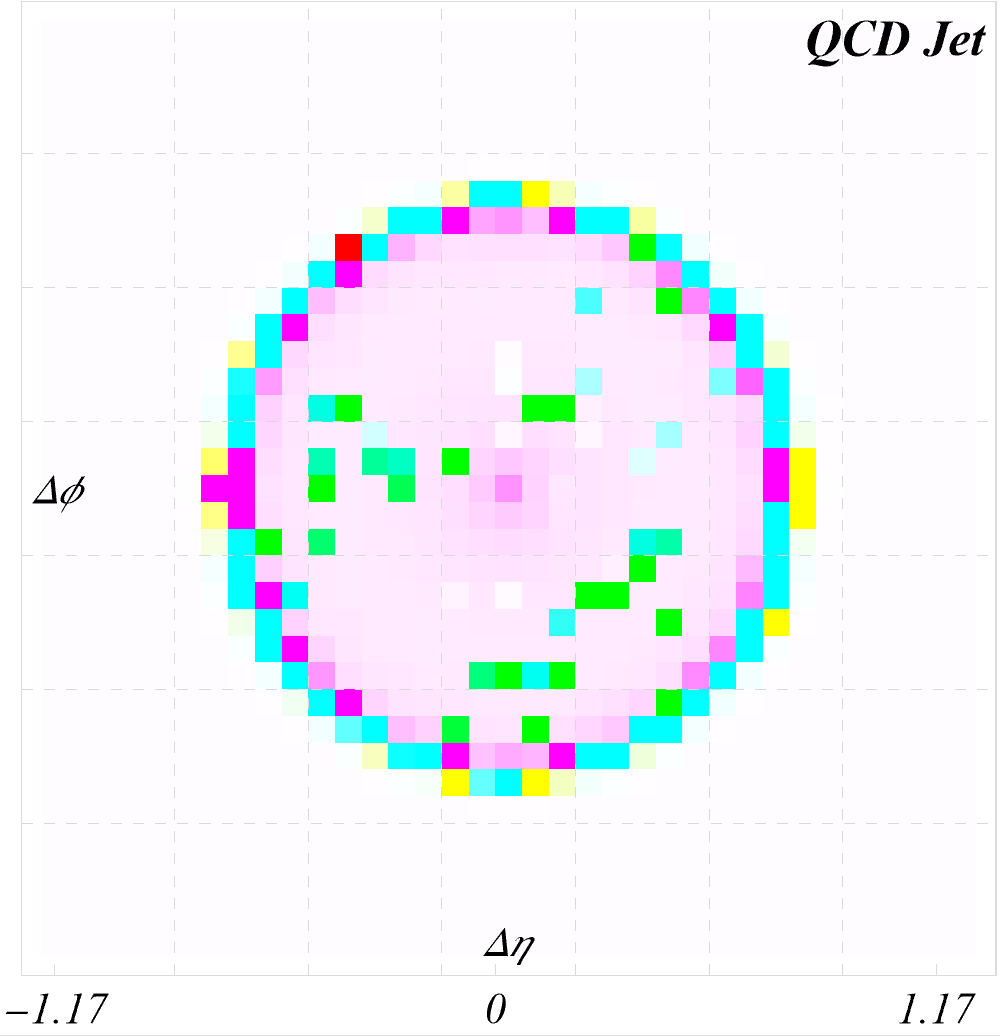}
        \caption{$35\times 35$, Standardized}
        \label{fig:qcdjetstandard}
    \end{subfigure}~
    \begin{subfigure}[t]{0.24\textwidth}
        \centering
        \includegraphics[height=4cm]{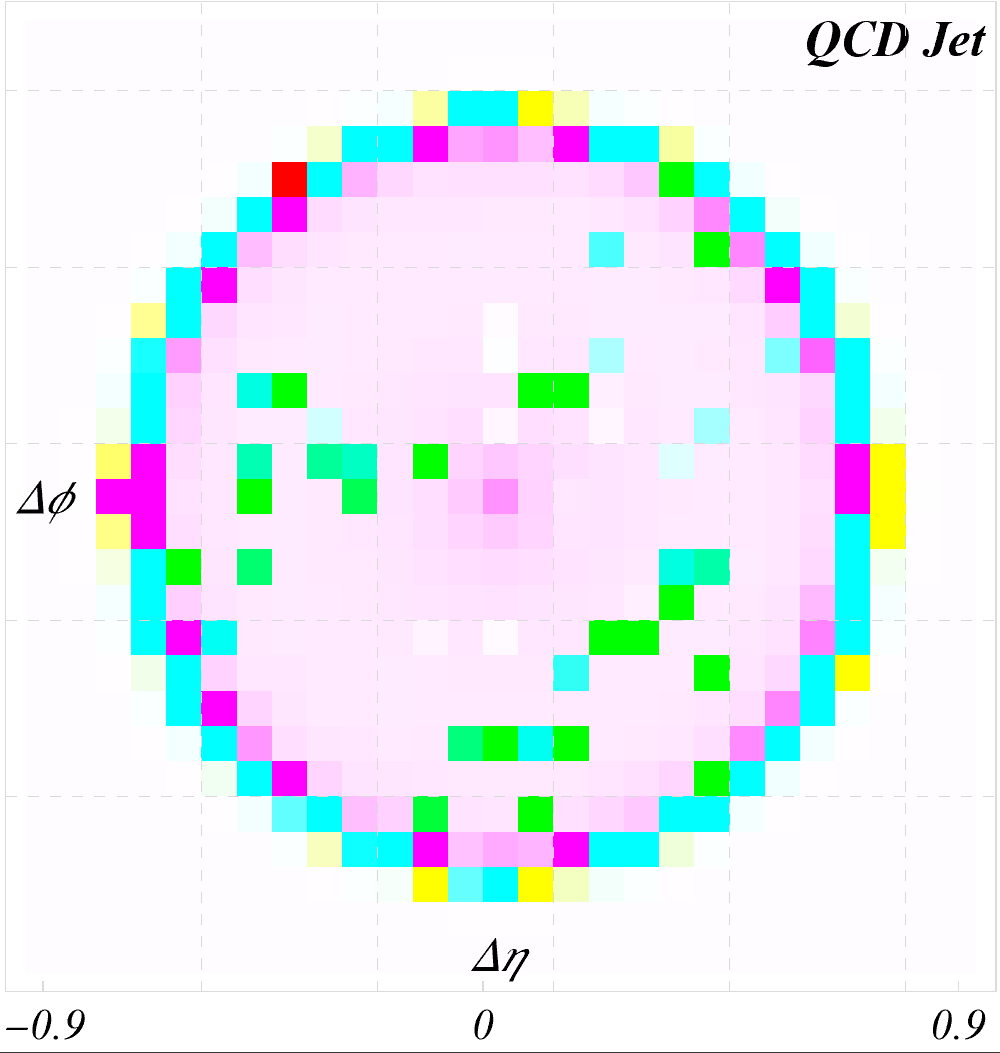}
        \caption{$35 \xrightarrow{Crop} 28$, Standardized}
        \label{fig:qcdjet32c28}
    \end{subfigure}~ 
    \begin{subfigure}[t]{0.24\textwidth}
        \centering
        \includegraphics[height=4cm]{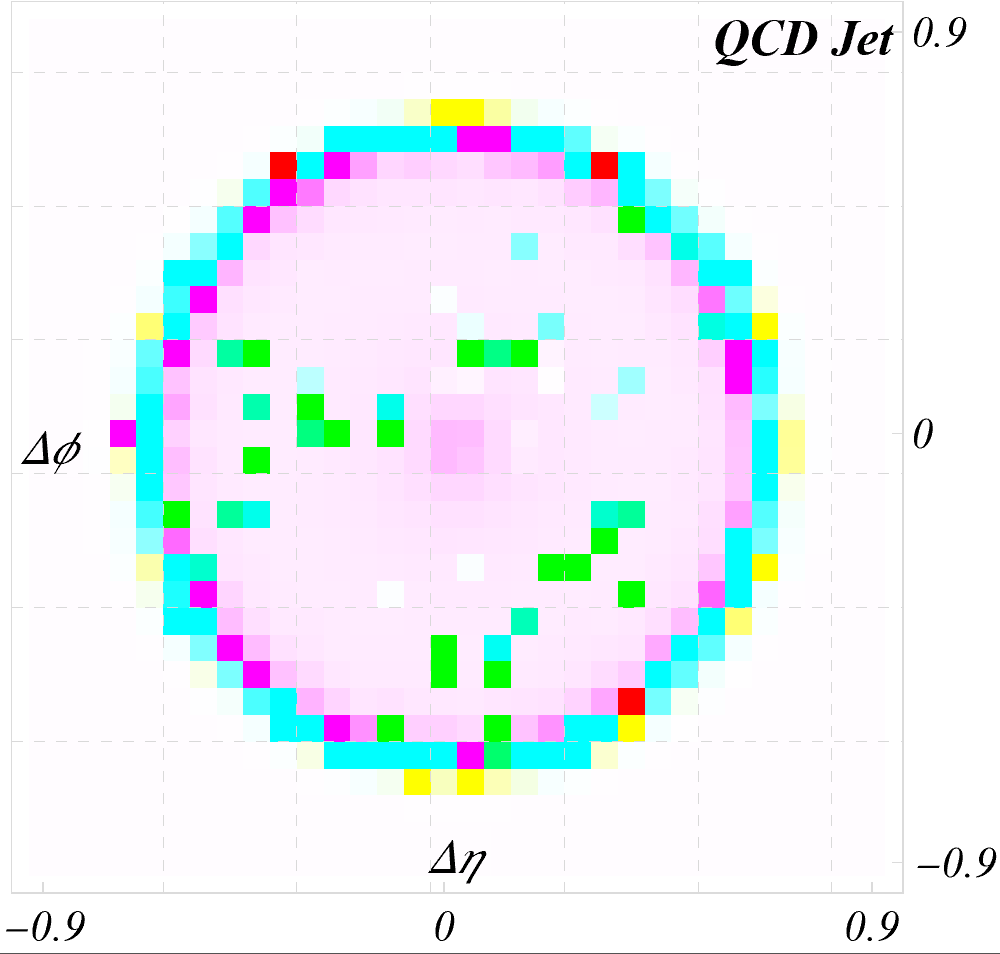}
        \caption{$40 \xrightarrow{Crop} 32$, Standardized}
        \label{fig:qcdjet40c32}
    \end{subfigure}
    \caption{\small Representative images of two jets for different resolutions and pixel standardization. Although images in our dataset are rendered in \texttt{RGB}, we present them here using the \texttt{CEI 1931 XYZ} color scheme \cite{xyzcolorCEI:1931}, for better visual understanding. All images from the top panel are of one representative top jet, and those below are of one QCD jet. The first column shows the jet constituents as a $35\times 35$ image in the $\Delta\eta - \Delta\phi$ plane, as the channels are combined and rendered. The second panel in the same row contains the same image, with the pixels standardized (the mean image of the dataset is subtracted from the original and then divided by the standard deviation image). The third column displays the images in the second, but cropped to $28\times 28$. The fourth column contains the same jets, first rendered in $40\times 40$, then cropped to $32\times 32$.}
    \label{fig:jetimages}
\end{figure*}
\begin{figure*}
    \centering
    \includegraphics[width=\textwidth]{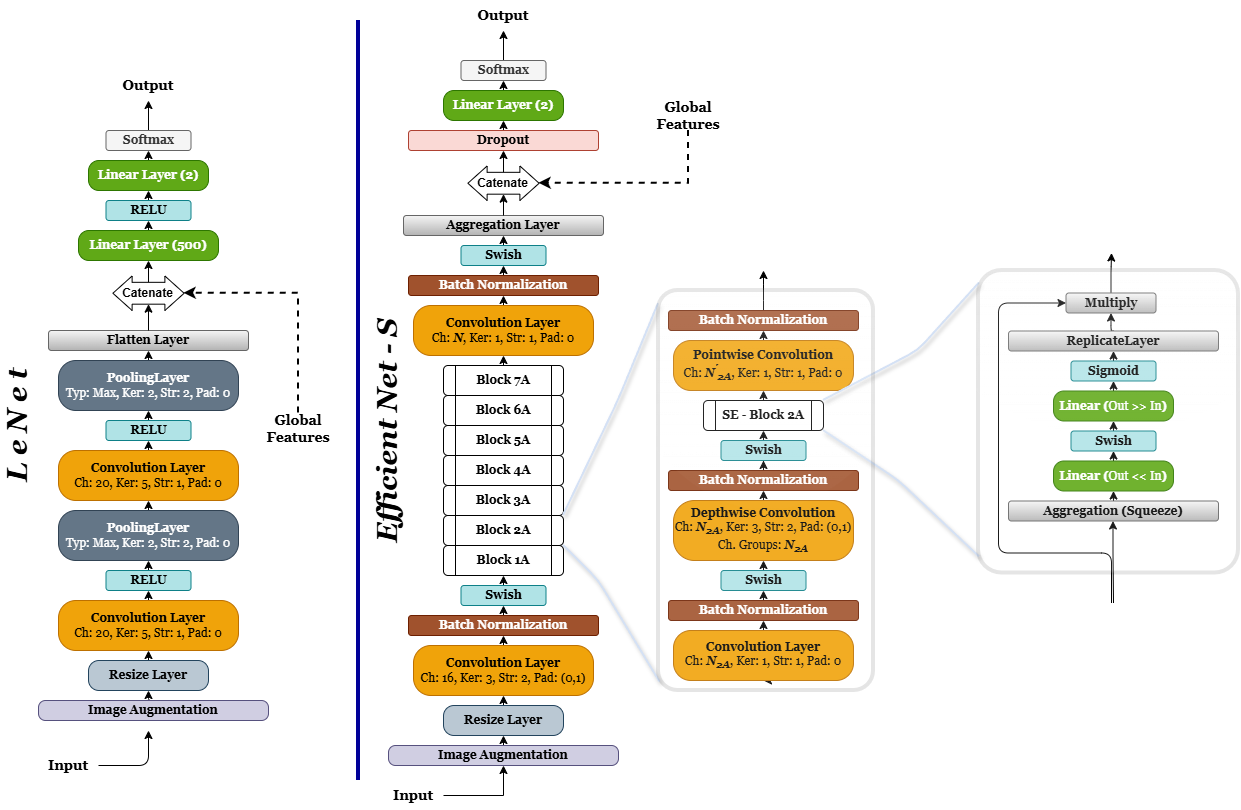}
    \caption{\small Schematics of network architectures used in this work. On the left is \textit{LeNet}, used as the benchmark CNN model. On the Right is the simplest Efficient Net (small) used. Representative schematics of the \textit{M-Conv} blocks and the squeeze and expand residual blocks within those are also shown.}
    \label{fig:netdiag}
\end{figure*}

\section{Methodology}
\label{sec:method}
\subsection{Machine Configuration}
\label{sec:machine}
All training and measurements, except for the measurement of inference speed, are performed on a single desktop PC with $64 GB$ RAM, a $13^{th}$ generation \textit{Intel\textsuperscript{\textcopyright} i9} processor, and a $12 GB$ \textit{NVIDIA\textsuperscript{\textcopyright} RTX A2000} graphics processor. The entire analysis is conducted in \textit{Mathematica}\textsuperscript{\textregistered} with the \textit{Wolfram Language}, which uses the \textit{Apache MXNet\textsuperscript{\textregistered}} framework in the backend. The original EfficientNet architectures (before modification) were taken from the \textit{Wolfram Net Repository} \cite{effnet:Wolfram}. To compare the inference speed of CNNs in this work with some external models, inference times were calculated on another machine with an $11 GB$ \textit{NVIDIA\textsuperscript{\textcopyright} GeForce GTX 1080-Ti} graphics processor.

\subsection{\textit{Data}: Images from Constituents}
\label{sec:data_images}
We have used a pre-existing dataset\footnote{\url{https://syncandshare.desy.de/public.php/dav/files/llbX3zpLhazgPJ6/?accept=zip}} that contains both top-quark and mixed light-quark-gluon jets. The objective, as mentioned before, is to identify top quark jets as signals and tag them as fat jets amid the sea of other jets. The events were generated with 14 TeV center-of-mass energy using Pythia8 \cite{Sjostrand:2014zea} followed by detector simulation performed using Delphes \cite{deFavereau:2013fsa}. The jets were constructed using the anti-kT algorithm \cite{Cacciari:2008gp} with FastJet \cite{Cacciari:2011ma}. The pseudorapidity $|\eta_j|$ is required to be $< 2$ for the fat top quark jet, whereas the radius parameter is chosen to be $0.8$. It is ensured that a matched parton-level top exists within $\Delta R=0.8$ of this signal jet. Moreover, all decay products of the top quark are required to be within $\Delta R=0.8$ of the jet axis. 4-momenta information is stored for the leading $200$ jet constituents. The leading top-quark jets lie within a $p_T$ range of $550 - 650$ GeV. The total sample consists of one million top jets alongside another million non-top jets, which are treated here as background. The dataset is divided into training / validation / test ($1.2 M/0.4 M/0.4 M$) sets.

We pre-process the data to create 3-channel images with transverse momentum ($p_T$), mass ($m$), and energy ($E$) in each channel, respectively. To create each channel from the jet constituents, the hardest constituent is considered to be the center of the channel, and the rest of the image is constructed by binning the remaining constituents in two dimensions: the difference in pseudo-rapidity ($\Delta\eta$) and that in the azimuthal angle ($\Delta\phi$). The paper associated with the original dataset \cite{Kasieczka:2019dbj} assumes the calorimeter resolution to be $0.04\times 2.25^\circ$, where a fat jet with a radius parameter $R=0.8$ can be covered with $40\times40$ pixels. Both $\Delta\eta$ and $\Delta\phi$ axes are retained from $-1.2$ to $1.2$ for this dataset, without the loss of any constituents. We keep more of the field of images than is usual practice in the community, where final images are generally cropped to $R=0.8$ \cite{Kasieczka:2017nvn,Kasieczka:2019dbj,Macaluso:2018tck}; our images are center-cropped at $R = 0.96$. To keep the images as small as possible, we use two sizes in the course of this work: $35\times 35$ (see Figs (\ref{fig:topjetorig}) and (\ref{fig:qcdjetorig}) ) and $40 \times 40$. In the original paper, the axis of the image is rotated in such a way that the second hardest $p_T$ object falls on the positive $\Delta\phi$ axis and the third hardest constituent in the first quadrant. We have found out that this results in a lower accuracy for classification; rather, we use an image augmentation layer to randomly flip the image along either axis or both during training. We do not use random cropping, as the jets are constructed objects and can---for all practical applications---be centered around the hardest $p_T$ constituent. Furthermore, to use \textit{N-subjettiness} later in our analysis, we also require each jet to have at least 4 constituents, which drops a meager 33 jets from the $2018 K$ dataset.

Standardization of input data has become common in ML, as it has been found to improve the performance of generic neural networks. We perform this standardization by creating a `mean' and a `standard deviation' image using a large number of random samples from the dataset. We respectively subtract from and divide every sample by the values of these arrays. This clearly isolates a fairly circular region in the $\Delta\eta-\Delta\phi$ plane containing the bulk of the jet structure (see Figs (\ref{fig:topjetstandard}) and (\ref{fig:qcdjetstandard}) ). Thus, we apply a center-crop to the $35\times 35$ images, making them $28\times 28$ ($R\sim 0.96$, see Figs (\ref{fig:topjet32c28}) and (\ref{fig:qcdjet32c28}) ). A similar size of the central region is obtained for the $40\times 40$ images when they are cropped to $32\times 32$ (see Fig.s (\ref{fig:topjet40c32}) and (\ref{fig:qcdjet40c32}) ). 


\subsection{\textit{Data}: Jet-Level Information}
\label{sec:data_global}
We have combined some global features of the jet to ascertain how much they improve the tagging efficiency of the CNNs. The features used in this work are inspired by refs \cite{ATL-PHYS-PUB-2021-028, Bhattacherjee:2022gjq}. The original dataset provides the four-momentum of up to 200 constituents per jet, from which all global features can be computed. For this purpose, we employ the \texttt{fastjet} package along with \texttt{EnergyCorrelator}  contrib from \texttt{fastjet-contrib-1.3.2}. 

In addition to the global four-momentum of the jet, our feature set includes the N-subjettiness variables \cite{Thaler:2010tr}, as well as several series of observables constructed from Energy Correlation Functions \cite{Larkoski:2013eya} and generalized correlator functions \cite{Moult:2016cvt}. The complete list of variables is as follows:

\begin{itemize}
    \item Jet four-momentum: $p_T$, $\eta$, $\phi$, $m_{\text{jet}}$
    \item Number of constituents: $N_{\text{cons}}$
    \item $N$-subjettiness ratios: $\tau_{21}^{\beta}$, $\tau_{32}^{\beta}$, $\tau_{43}^{\beta}$, $\tau_{54}^{\beta}$, $\tau_{65}^{\beta}$ for $\beta = 0.5$, $1.0$, $2.0$
    \item $C$-series: $C_1^\beta$, $C_2^\beta$, $C_3^\beta$ for $\beta = 1.0,\,2.0$
    \item $D$-series\footnote{The parameter choices for the D-series features follow \cite{Larkoski:2014gra, Larkoski:2015kga, Larkoski:2014zma}.}: $D_2^{\alpha,\beta}$ for \(\alpha, \beta = 1.0,\,2.0\), and the four \(D_3\) variables defined in \cite{Bhattacherjee:2022gjq}
    \item U-series: \(U_1^\beta, U_2^\beta, U_3^\beta\) for \(\beta = 0.5,\,1.0,\,2.0\)
    \item M-series: \(M_2^\beta, M_3^\beta\) for \(\beta = 1.0,\,2.0\)
    \item N-series: \(N_2^\beta, N_3^\beta\) for \(\beta = 1.0,\,2.0\)
    \item L-series: \(L_2, L_3\) as defined in \cite{ATL-PHYS-PUB-2021-028}
\end{itemize}

Certain global features, such as \(U_3^\beta\) and \(M_3^\beta\), require at least 4 constituents for their computation. As mentioned in the previous sub-section, we discard all jets with \(N_{\text{cons}} < 4\). 

The roles of the specialized series are summarized below:
\begin{itemize}
    \item \textbf{$N$-subjettiness ($\tau_{N}$):} Measures the alignment of jet radiation along $N$ candidate axes. We use ratios($\tau_N/\tau_{N-1}$) up to $\tau_{65}$ with angular exponents $\beta \in \{0.5, 1.0, 2.0\}$. These variables quantify the $N$-prong nature of the jet.
    \item \textbf{C-Series ($C_{n}^{\beta}$):} Identifies the number of hard energy centers (prongs) within a jet by correlating the energy deposits of $n$ particles.
    \item \textbf{D-Series ($D_{2}, D_{3}$):} Specifically engineered ratios of energy correlation functions that provide optimal discrimination between different prong counts, with $D_{3}$ tuned for the 3-prong top-quark signature.
    \item \textbf{U, M, and N Series:} Generalized energy correlators that provide granular information on the angular distribution and energy flow of the jet constituents.
    \item \textbf{L-Series ($L_{2}, L_{3}$):} Shape observables that complement the $N$-subjettiness variables to improve the resolution of multi-prong topologies.
\end{itemize}
For complete definitions and a detailed description of these variables, we refer the reader to refs \cite{Bhattacherjee:2022gjq,ATL-PHYS-PUB-2021-028}.

\subsection{Model Architecture}
\label{sec:mlmodels}
Historically, one of the most successful convolutional neural networks (CNNs), \textit{LeNet} \cite{lenetPaper}, is also the oldest and one of the simplest. \textit{LeNet-5}, with alternating convolutional and pooling layers, has been the dependable standard for smaller images (see the left side of Fig. (\ref{fig:netdiag}) for its architecture). Over the decades following \textit{LeNet}, CNNs continued to scale up to classify images at increasingly higher resolutions. AlexNet \cite{AlexNetPaper}, Inception nets \cite{InceptionV1, InceptionV2V3}, ResNets \cite{ResNet1} and their wider versions \cite{wideResNet}, DenseNets \cite{DenseNet}, and MobileNets \cite{MobileNetV1, MobileNetV2, MobileNetV3, MobileNetV4} are only a few examples of these. To go deeper and wider, these newer architectures used varied techniques, such as `residual' connections to feed information from earlier layers to the latter parts of the network, point-wise $(1 \times 1)$ and depthwise convolutions for dimension reduction to remove computational bottlenecks, batch normalization, and mixtures of these methods \cite{InceptionV4, ResNext}. Targeting general-purpose image classification, these models were optimized for the ImageNet dataset \cite{ImageNet2019,ImageNet2021} and its equivalents; thus, computationally, they are quite expensive.

\begin{table}[!tb]
    \centering
    \renewcommand{\arraystretch}{1.5}
    \begin{tabular}{|c|c|c|}
        \hline
        Layers           & \multicolumn{2}{c|}{Input Resolution} \\ \cline{2-3}
                         & $3\times 40\times 40$  & $3\times 64\times 64$ \\
        \hline
        Conv $3\times 3$ & $16\times 20\times 20$ & $16\times 32\times 32$ \\
        Block 1A         & $8\times 20\times 20$  & $8\times 32\times 32$ \\
        Block 2A         & $8\times 10\times 10$  & $16\times 16\times 16$ \\
        Block 3A         & $16\times 5\times 5$   & $16\times 8\times 8$ \\
        Block 4A         & $24\times 3\times 3$   & $32\times 4\times 4$ \\
        Block 5A         & $32\times 3\times 3$   & $48\times 4\times 4$ \\
        Block 6A         & $56\times 2\times 2$   & $80\times 2\times 2$ \\
        Block 7A         & $96\times 2\times 2$   & $136\times 2\times 2$ \\
        Conv $3\times 3$ & $400\times 2\times 2$  & $536\times 2\times 2$ \\
        Aggregate        & $400$                  & $536$ \\
        FC \& Out        & $2$                    & $2$  \\
        \hline
    \end{tabular}
    \caption{\small \textit{Effnet-S} configurations for two different input resolutions. For more details, see sec. \ref{sec:mlmodels}.}
    \label{tab:effnetConfig}
\end{table}

\begin{table*}[!tb]
    \centering
    \renewcommand{\arraystretch}{1.25}
    \begin{tabular}{|c|c|c|c|c|c:c|c:c|c:c|}
    \hline\hline
        \textbf{Network}        & \textbf{Input}         & \textbf{Img. Res.}      & \textbf{No. of}     &  \textbf{Inference}    & \multicolumn{2}{c|}{\textbf{Accuracy}}  & \multicolumn{2}{c|}{\textbf{AUC}}   & \multicolumn{2}{c|}{\textbf{Bkg. Rejection}} \\ 
        \textbf{Type}           & \textbf{Array}          & ($Org\to$ & \textbf{Param.s} & \textbf{Time} ($ms$) & \multicolumn{2}{c|}{(\%)}  & \multicolumn{2}{c|}{ } & \multicolumn{2}{c|}{\textbf{at TPR}} \\ 
        \cline{6-11}
        & ($3\times n\times n$)          & $Crp$) & \textbf{Param.s} & /100 Jets & Avg.  & Best  & Avg.  & Best  & $50\%$   & $30\%$ \\
        \hline
                       & $28$ & $35\to 28$ & $428 K$  & $5.7(8)$  & $92.42(3)$ & $92.48$ & $97.74(1)$ & $97.76$ & $156(2)$ & $480(18)$ \\
 \textit{\large\underline{LeNet}} 
                       & $32$ &            & $653 K$  & $6.1(9)$  & $92.40(2)$ & $92.43$ & $97.74(1)$ & $97.74$ & $154(3)$  & $483(13)$ \\
        (Images)     & $40$ &            & $1253 K$ & $6.4(8)$ & $92.45(3)$ & $\mathbf{92.50}$ &
                       $97.76(2)$ & $\mathbf{97.78}$ & $\mathbf{161(4)}$  & $\mathbf{514(23)}$ \\
                       \cdashline{2-11}
                       & $32$ & $40\to 32$ & $653 K$  & $11.9(10)$  & $92.41(4)$ & $92.48$ & $97.75(1)$ & $97.76$ & $161(3)$  & $512(25)$ \\
                       & $40$ &            & $1253 K$ & $13.4(18)$  & $92.42(5)$ & $92.48$ & $97.74(1)$ & $97.74$ & $154(3)$  & $476(11)$ \\
                       \hdashline
                       & $28$ & $35\to 28$ & $485 K$  & $5.7(8)$  & $93.33(2)$ & $93.37$ & $98.24(0)$ & $98.24$ & $258(5)$  & $916(25)$ \\
        (Images        & $32$ &            & $710 K$  & $6.1(8)$  & $93.34(2)$ & $\mathbf{93.38}$ &                    $98.24(1)$ & $\mathbf{98.26}$ & $257(3)$  & $942(31)$ \\
             +         & $40$ &            & $1310 K$ & $6.6(9)$  & $93.31(3)$ & $93.36$ &               $98.23(1)$ & $98.25$ & $255(5)$  & $909(34)$ \\
                      \cdashline{2-11}
        Global         & $32$ & $40\to 32$ & $710 K$  & $14.5(16)$  & $93.33(2)$ & $93.36$ &                    $98.24(0)$ & $98.24$ & $255(2)$  & $\mathbf{951(43)}$ \\
        Feat.s)        & $40$ &            & $1310 K$ & $14.8(7)$  & $93.34(2)$ & $93.37$ &                    $98.24(1)$ & $98.25$ & $\mathbf{263(3)}$  & $906(32)$ \\
                       \hline\hline
                       & $36$ (\textit{S0}) & $35\to 28$ & $158 K$ & $7.3(9)$  & $92.31(11)$ & $92.50$ & $97.73(4)$ & $97.80$ & $160(5)$  & $536(24)$ \\
   \textit{\large\underline{EffNet-S}} 
                       & $40$ (\textit{S1}) &            & $184 K$ & $8.0(7)$  & $92.35(8)$ & $92.46$ & $97.75(3)$ & $97.76$ & $163(3)$  & $552(18)$ \\
                       & $48$ (\textit{S2}) &            & $215 K$ & $8.7(8)$  & $92.39(9)$ & $92.53$ & $97.76(4)$ & $97.80$ & $\mathbf{167(5)}$  & $569(30)$ \\
                       & $56$ (\textit{S3}) &            & $258 K$ & $9.6(8)$  & $92.58(1)$ & $92.72$ & $97.85(3)$ & $97.90$ & $171(6)$  & $\mathbf{585(32)}$ \\
         (Images)      & $64$ (\textit{S4}) &            & $305 K$ & $10.6(8)$ & $92.20(40)$ & $\mathbf{92.75}$ &                    $97.74(11)$ & $\mathbf{97.92}$ & $156(16)$ & $499(78)$ \\
                       \cdashline{2-11}
                       & $36$ (\textit{S0}) & $40\to 32$ & $158 K$ & $13.5(10)$  & $91.40(60)$ & $91.91$ & $97.37(13)$ & $97.50$ & $127(11)$  & $431(49)$ \\
                       & $40$ (\textit{S1}) &            & $184 K$ & $13.9(10)$  & $90.5(12)$ & $91.80$ & $97.22(26)$ & $97.39$ & $112(15)$  & $361(57)$ \\
                       & $48$ (\textit{S2}) &            & $215 K$ & $14.8(10)$  & $91.4(10)$ & $92.05$ & $97.35(23)$ & $97.56$ & $126(18)$  & $407(64)$ \\
                       & $56$ (\textit{S3}) &            & $258 K$ & $15.8(7)$  & $92.0(4)$ & $92.31$ & $97.60(10)$ & $97.68$ & $143(9)$  & $460(45)$ \\
                       & $64$ (\textit{S4}) &            & $305 K$ & $16.5(9)$  & $91.9(6)$ & $92.42$ & $97.52(18)$ & $97.74$ & $137(17)$  & $434(63)$ \\
                       \hdashline
                       & $36$ (\textit{S0}) & $35\to 28$ & $242 K$ & $8.2(9)$  & $93.28(2)$ & $93.32$ & $98.21(1)$ & $\mathbf{98.23}$ & $242(5)$  & $890(34)$ \\
       (Img.s +        & $40$ (\textit{S1}) &            & $276 K$ & $7.9(9)$  & $93.26(2)$ & $93.30$ &                      $98.21(1)$ & $98.22$ & $241(7)$  & $859(34)$ \\
       Glob. Feat.s)   & $48$ (\textit{S2}) &            & $315 K$ & $8.4(9)$  & $93.20(2)$ & $93.30$ &                      $98.18(2)$ & $98.20$ & $241(10)$  & $839(51)$ \\
                       & $56$ (\textit{S3}) &            & $368 K$ & $10.0(10)$  & $93.31(5)$ & $\mathbf{93.38}$ &      $98.21(1)$ & $98.22$ & $232(5)$  & $817(26)$ \\
                       & $64$ (\textit{S4}) &            & $422 K$ & $11.4(9)$  & $93.15(5)$ & $93.24$ & $98.11(3)$ & $98.15$ & $208(8)$  & $707(48)$ \\
                       \cdashline{2-11}
                       & $36$ (\textit{S0}) & $40\to 32$ & $242 K$ & $13.8(8)$  & $93.24(8)$ & $93.31$ & $98.20(4)$ & $\mathbf{98.23}$ & $244(13)$  & $\mathbf{897(74)}$ \\
                       & $40$ (\textit{S1}) &            & $276 K$ & $13.6(12)$  & $93.26(2)$ & $93.31$ & $98.20(1)$ & $98.21$ & $\mathbf{247(6)}$  & $868(25)$ \\
                       & $48$ (\textit{S2}) &            & $315 K$ & $14.4(9)$  & $93.20(3)$ & $93.26$ & $98.17(2)$ & $98.20$ & $235(6)$  & $864(30)$ \\
                       & $56$ (\textit{S3}) &            & $368 K$ & $15.7(9)$  & $93.22(9)$ & $93.35$ & $98.18(3)$ & $98.23$ & $237(9)$  & $829(44)$ \\
                       & $64$ (\textit{S4}) &            & $422 K$ & $17.0(10)$  & $93.21(5)$ & $93.27$ & $98.15(3)$ & $98.18$ & $225(13)$  & $778(71)$ \\
                       \hline\hline
        DeepTop & $1\times$ & $60\to 40$ & & & & & & & & \\
        \cite{Macaluso:2018tck,Kasieczka:2017nvn,Kasieczka:2019dbj}
                       & $37\times 37$ & $\to 37$ & $610 K$  & -- & -- & $93.00$ & -- & $98.10$ & -- & $914(14)$ \\ \hdashline
                    \multicolumn{11}{|c|}{ResNeXt-50 \cite{ResNext}} \\ \hdashline
    Ref. \cite{Kasieczka:2019dbj} & $1\times$ & & & & & & & & & \\
                       & $64\times 64$ & $96\to 64$ & $1460 K$  & -- & -- & $93.60$ & -- & $98.40$ & -- & $1122(47)$ \\
    Ref. \cite{Qu:2019gqs} & $1\times$ & & & & & & & & & \\
                       & $64\times 64$ & $96\to 64$ & $1460 K$  & 22 & -- & $93.60$ & -- & $98.37$ & $302(5)$ & $1147(58)$ \\ \hline
    \end{tabular}
    \caption{\small Single number performance metrics for ML models performed on the total test dataset. From the second to the seventh column, we respectively report the input array sizes, the resolution of the generated jet images ($A\to B$ means the original image was $A\times A$, and it was cropped to $B\times B$), total number of parameters in the network, inference time for 100 jets in milliseconds, overall accuracy, and area under the ROC curve (AUC). The last two columns list the QCD background rejection at top tagging efficiencies of $50\%$ and $30\%$, respectively. The values in bold-face represent the best metrics for a specific type of Network.}
    \label{tab:pointres}
\end{table*}

Though scaling CNNs according to image resolution has been attempted before \cite{long2015fully}, EfficientNets \cite{EffNetV1, EffNetV2} were the first to solve the scaling problem by using `compound scaling', connecting network \textit{depth} and \textit{width} to image \textit{resolution}. Given a baseline network (found from a neural architecture search, or NAS), these quantities scale as:

\begin{align}
\begin{alignedat}{2}
&\text{depth coefficient } d &&= \alpha^\phi \\
&\text{width coefficient } w &&= \beta^\phi \\
&\text{resolution coefficient } r &&= \gamma^\phi
\end{alignedat}
\end{align}

\begin{figure*}[t!]
    \centering
    \begin{subfigure}[t]{0.475\textwidth}
        \centering
        \includegraphics[height=5.25cm]{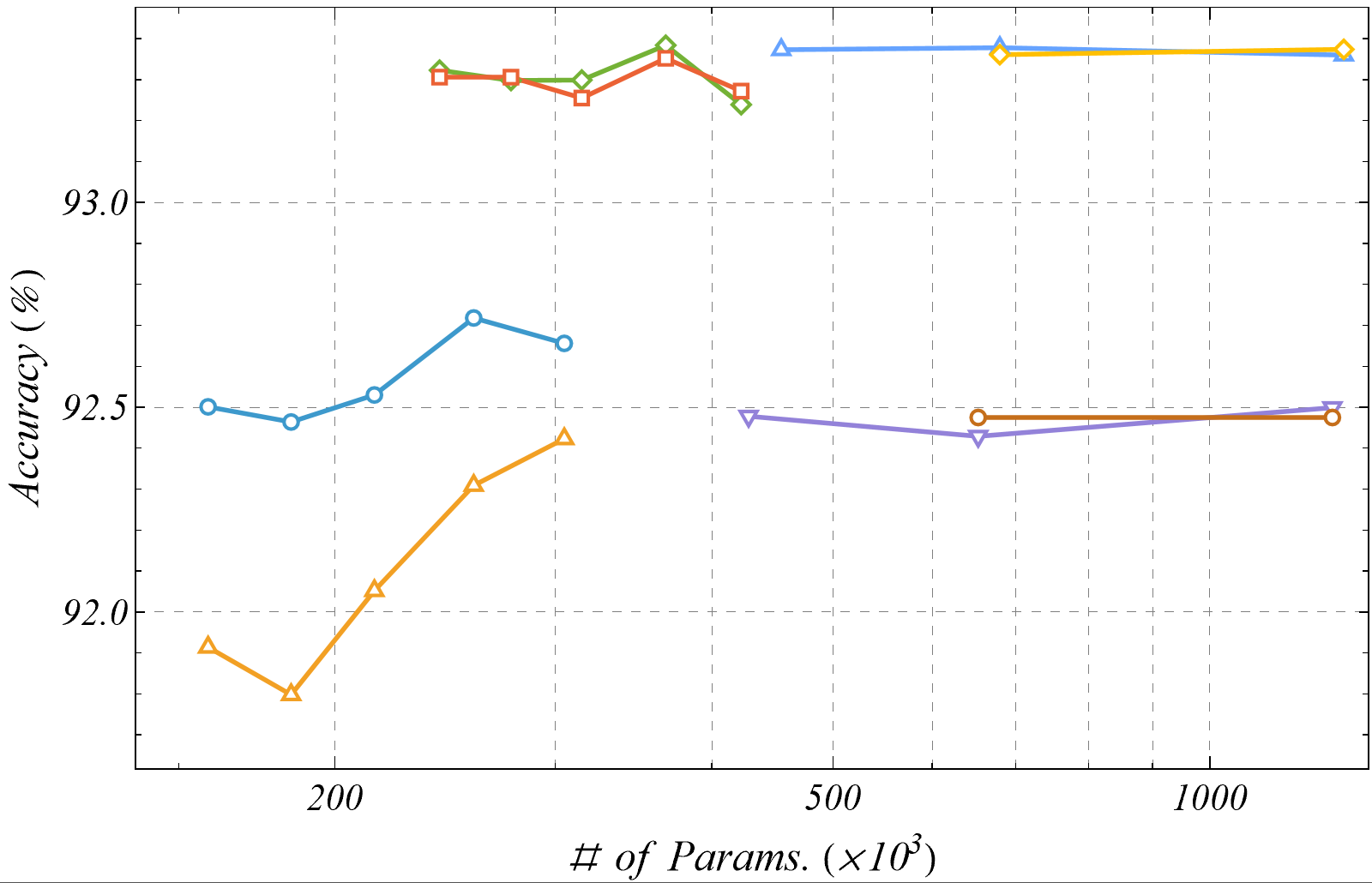}
        \caption{Model Complexity vs Best Accuracy}
        \label{fig:compbest}
    \end{subfigure}~ 
    \begin{subfigure}[t]{0.475\textwidth}
        \centering
        \includegraphics[height=5.25cm]{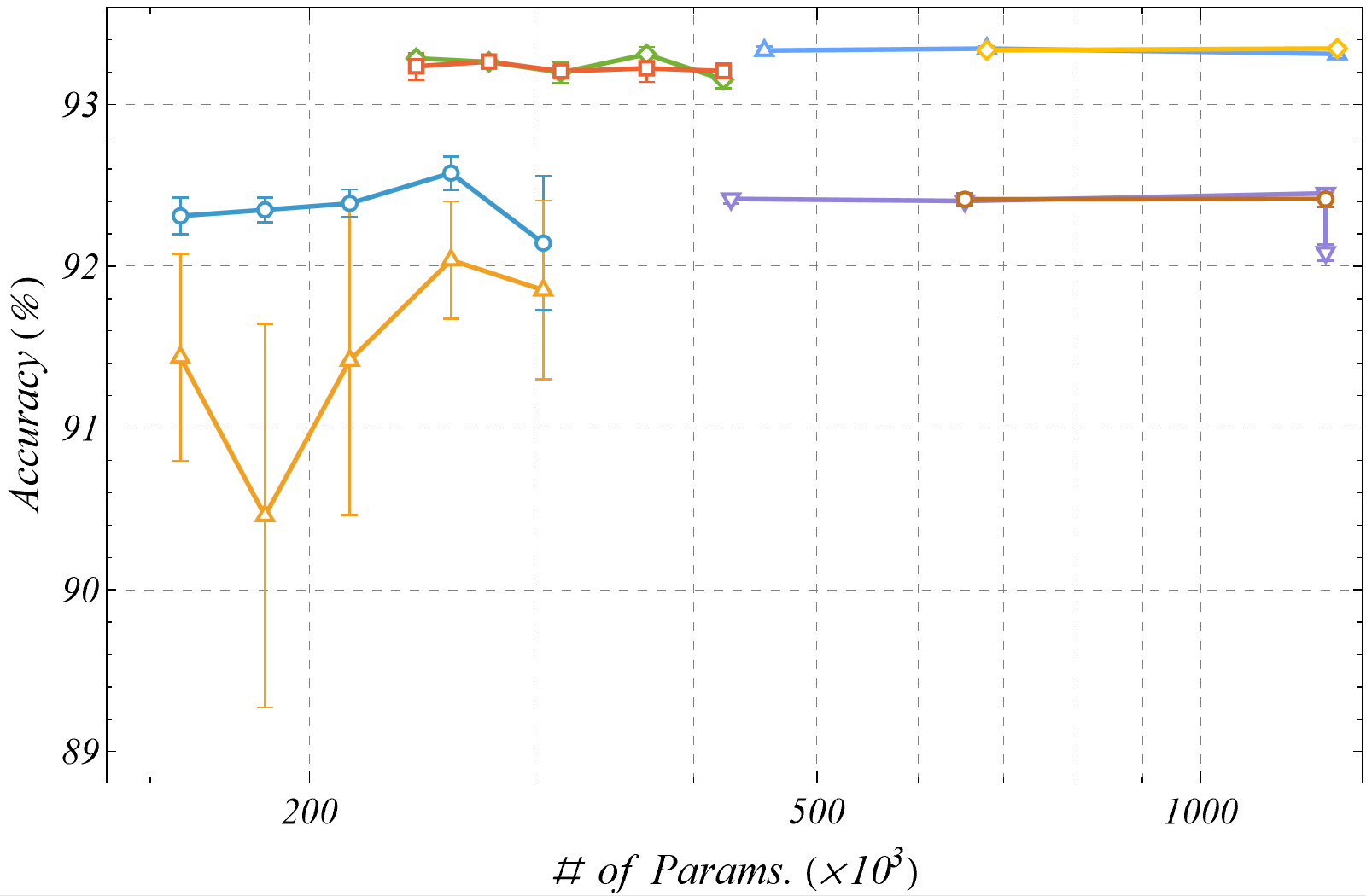}
        \caption{Model Complexity vs Accuracy Dist.}
        \label{fig:compdist}
    \end{subfigure}\\
    \begin{subfigure}[t]{0.25\textwidth}
        \centering
        \includegraphics[height=6.3cm]{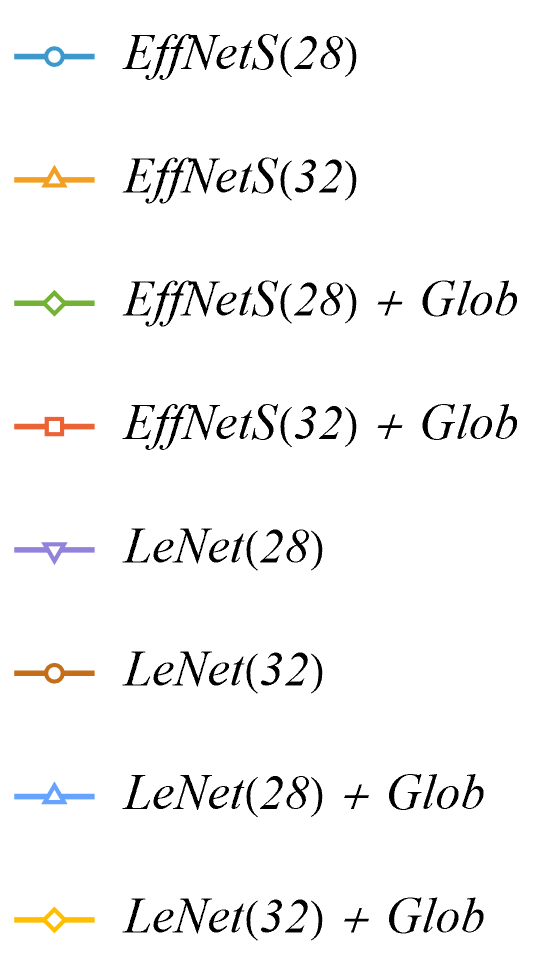}
        \caption{Plot Legend}
        \label{fig:complegend}
    \end{subfigure}~ 
    \begin{subfigure}[t]{0.5\textwidth}
        \centering
        \includegraphics[height=6.3cm]{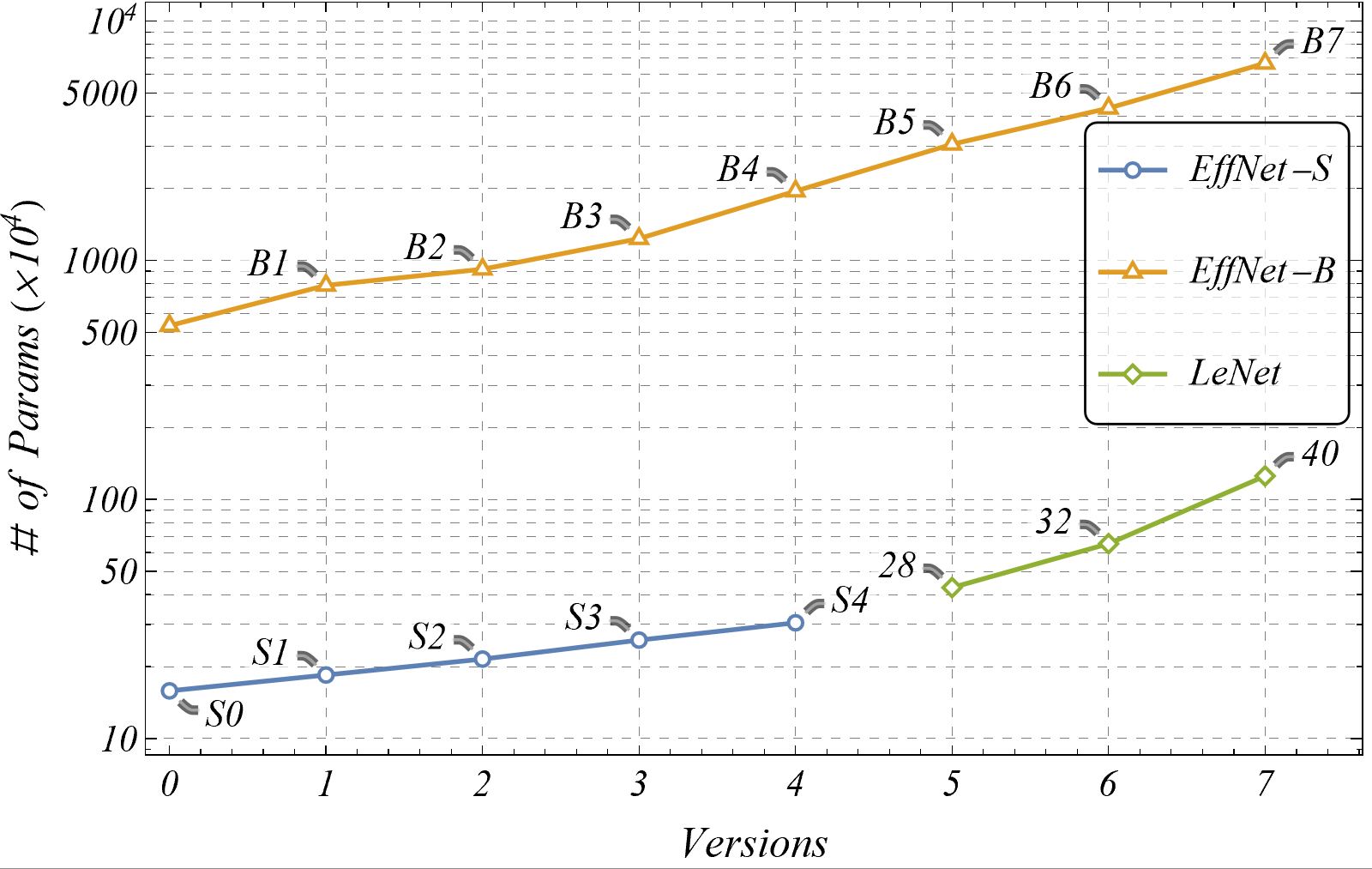}
        \caption{Size Comparison between \textit{EffNet-B}, \textit{EffNet-S}, and \textit{LeNet}}
        \label{fig:compall}
    \end{subfigure}
    \caption{\small Performance comparison of different CNN architectures: fig. (\textit{a}) shows the best accuracies vs. number of model parameters of the nets used, fig. (\textit{b}) shows the median accuracies and error-bars in the same setting, fig. (\textit{c}) is the legend for (\textit{a}) and (\textit{b}), and fig. (\textit{d}) is a comparison of sizes of the networks compared to traditional \textit{EfficientNet} versions.}
    \label{fig:compnets}
\end{figure*}

where $\alpha\geq 1$, $\beta\geq 1$, $\gamma\geq 1$. Here, increasing $\phi$, the scaling parameter, increases the complexity of the network in a compounded manner. Noting that doubling the net-depth doubles the computational cost (floating point operations per second, or FLOPS), but doubling the other two quadruples the FLOPS, these also follow an empirical constraint: $\alpha\cdot\beta^2\cdot\gamma^2\simeq 2$, so that increasing $\phi$ by some $\phi_0$ increases the FLOPS by $2^{\phi_0}$. The authors of the first EfficientNet paper performed a grid search on $\alpha$, $\beta$, and $\gamma$ and obtained the values $1.2$, $1.1$, and $1.15$, respectively. Starting with $\phi=0$ and increasing $\phi$ by 1 each time, they obtained eight networks ($B_0$ to $B_7$) with increasing complexity. The base network ($B_0$) works on arrays of dimension $224\times 224$, while the largest of these models ($B_7$) takes in arrays of dimension $600\times 600$. There are seven main building blocks of an EfficientNet, each containing multiple parts. Building upon other successful lightweight models \cite{MobileNetV2,MNASnet}, EfficientNets replace a fully convolutional operation with a depthwise convolution (a single filter per channel) and a pointwise ($1\times 1$) convolution (see the blown up schematic of \textit{Block 2A} in Fig. (\ref{fig:netdiag}) ). This significantly reduces the computational cost while maintaining similar performance. When stacked together, the blocks mentioned above also perform a function known as inverted bottlenecking, embedding the input manifold into a lower-dimensional manifold. Each of these blocks also contains a \textit{squeeze and expansion} block \cite{squeeze}, with a residual connection maintaining the information flow, as well as a combination of \textit{Swish}, \textit{RELU}, and \textit{Sigmoid} activation functions. More details can be found in the original paper \cite{EffNetV1}. While the width coefficient scales up the channel dimension of each block, the depth coefficient determines the number of times each type of block is repeated, making the network deeper. These networks also scale up the dropout probability to regularize the larger models. All versions of EfficientNet outperform other CNNs that are much larger in size (in terms of accuracy) while being faster during inference ($B_0$, with $5.3~M$ parameters, achieves a top-5 accuracy of $93.3\%$ for ImageNet, compared to the widely used ResNet-50 ($26~M$ parameters)—which has a top-5 accuracy of $93.0\%$; the FLOPS for ResNet-50 is $11$ times that of EfficientNet-$B_0$).

We begin with \textit{LeNet} in this analysis as a benchmark for performance. Following its historical usage, the smallest resolution of jet images we use is $28\times 28$, increasing to $32\times 32$ and $40\times 40$. Images larger than these make the jet constituents sparser, and CNNs with even higher-resolution inputs are needed for their analysis. The network we focus on here is \textit{EfficientNet}. While they have been used on smaller images, such as the \texttt{CIFAR-10} and \texttt{CIFAR-100} datasets, these have been upsampled to fit the $224\times 224$ input dimensions \footnote{Author's answer: \url{https://github.com/tensorflow/tpu/issues/421}.}. This defeats one of the main purposes of this work (working with constrained computational resources), both in terms of RAM and VRAM. 

\begin{figure}
    \centering
    \includegraphics[width=\linewidth]{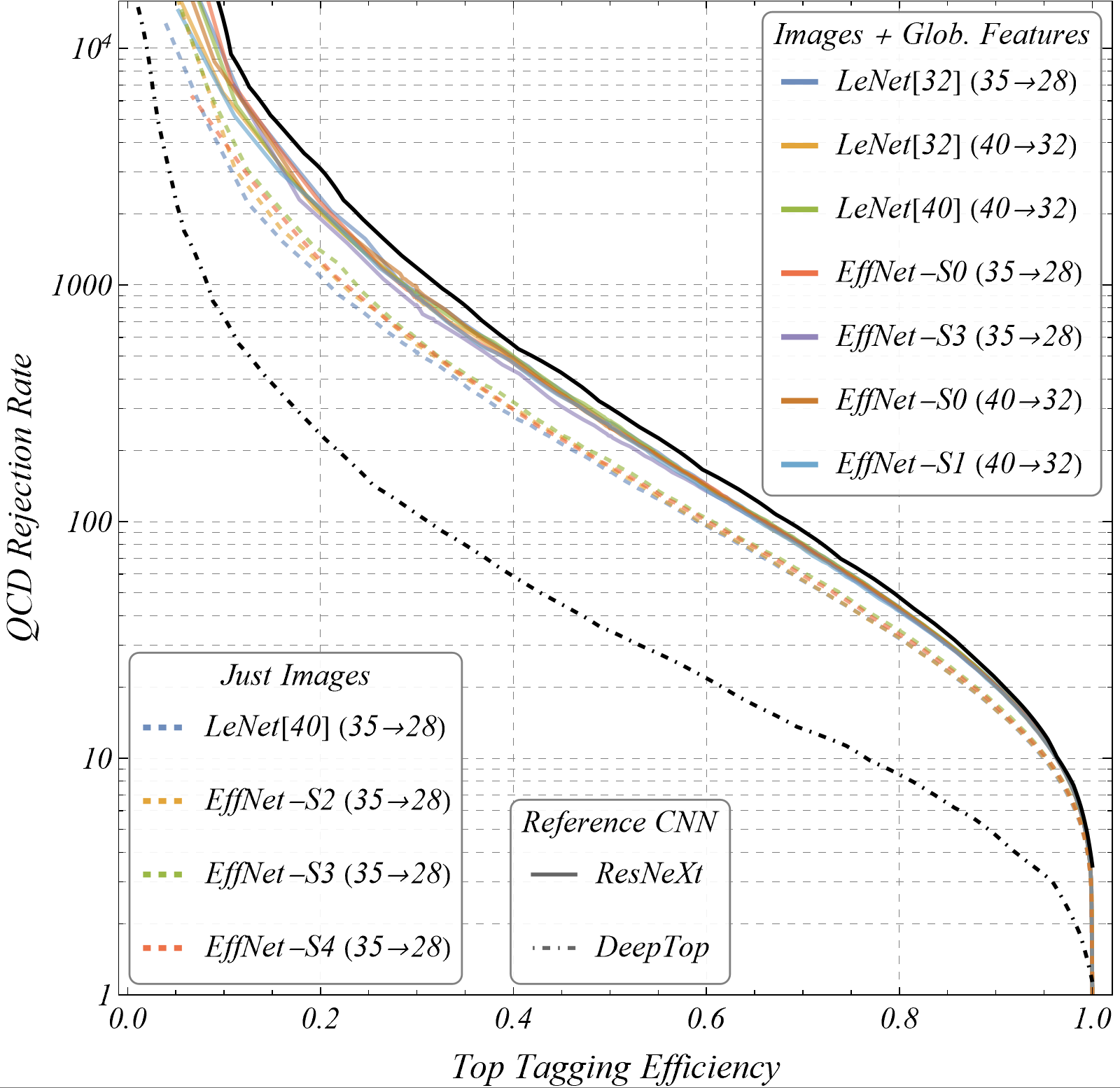}
    \caption{\small ROC curve (signal efficiency vs. background rejection) for some of the networks (with some metric highest among others; table (\ref{tab:pointres}) has them in bold-face). The solid lines are for models working on both images and global features, dashed ones correspond to those with only images as input. The black solid and dot-dashed curves show similar ROC curves for the \textit{ResNext-50} and the \textit{DeepTop} networks, respectively \cite{Kasieczka:2019dbj,Kasieczka:2017nvn}.  
    }
    \label{fig:roc}
\end{figure}

To circumvent this without changing the hyperparameters of the original EfficientNet architecture (and using the prescribed values of $\alpha$, $\beta$, and $\gamma$), we use negative $\phi$ values to scale down the input resolution. The smallest resolution thus achieved is $36\times 36$. Any smaller, and the bottlenecks within the network would not hold without modification. We choose the starting resolution to be $36\times 36$, corresponding to $\phi = -13$, then repeatedly increase $\phi$ by 1 to get the rest of the configurations.  We thus create 5 variants, the last one corresponding to $\phi = -9$\footnote{Similar efforts can be found here: \url{https://github.com/qubvel/efficientnet/issues/26}}. The other important parameter to set is the dropout probability, which is gradually increased from $B_0$ to $B_7$, as more regularization is needed for larger networks. We linearly extrapolated these values using the width and depth coefficients and found that for input array widths $<90$, we do not need dropouts. Altogether, we will denote this smaller architecture as EfficientNet-Small (\textit{EffNet-S}) for this work. For these values of $\phi$, every block appears only once in the smaller \textit{EffNet-S} structures. We have thus indexed the blocks `$i$-A' ($i=1,\,2,\,\ldots$), to keep the possibility of larger networks open\footnote{This raises an obvious question: as these CNNs were fine-tuned to detect specific \textit{shapes} (with detail) from larger images, unlike in our case—where the jet-images are smaller and are mere representations with a much larger importance to finer structures—could it be possible to systematically search for and create versions of scalable networks for this specific task, albeit smaller? Though intriguing, this question is outside the scope of the present work. However, glimpses of this possibility may be perceived from the present analysis.}. Table (\ref{tab:effnetConfig}) lists the input dimensions for this architecture for two different input resolutions: $40\times 40$ and $64\times 64$.

\begin{figure*}[t!]
    \centering
    \begin{subfigure}[t]{0.475\textwidth}
        \centering
        \includegraphics[height=5.25cm]{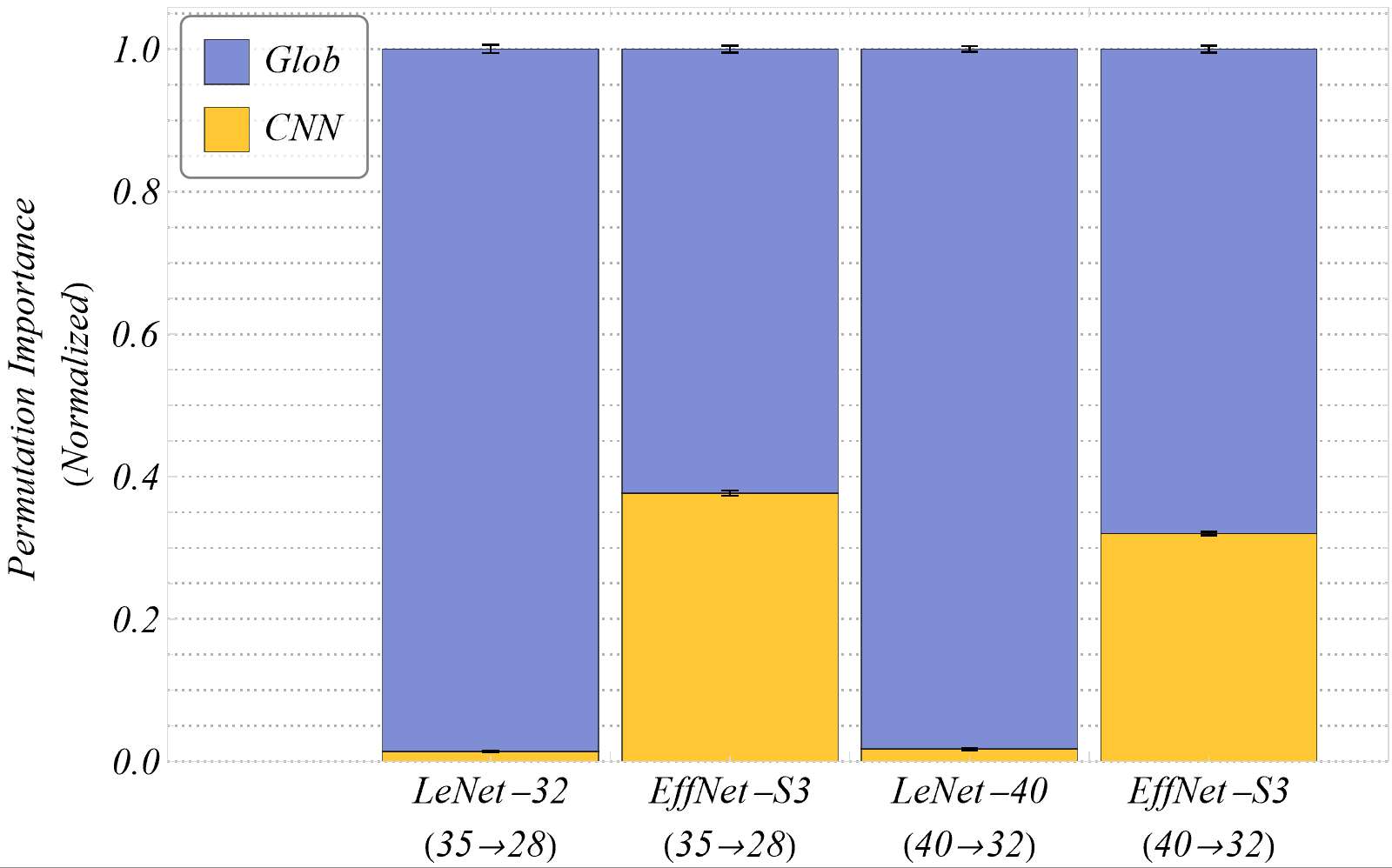}
        \caption{CNN vs All Global Features}
        \label{fig:importanceFull}
    \end{subfigure}~ 
    \begin{subfigure}[t]{0.475\textwidth}
        \centering
        \includegraphics[height=5.25cm]{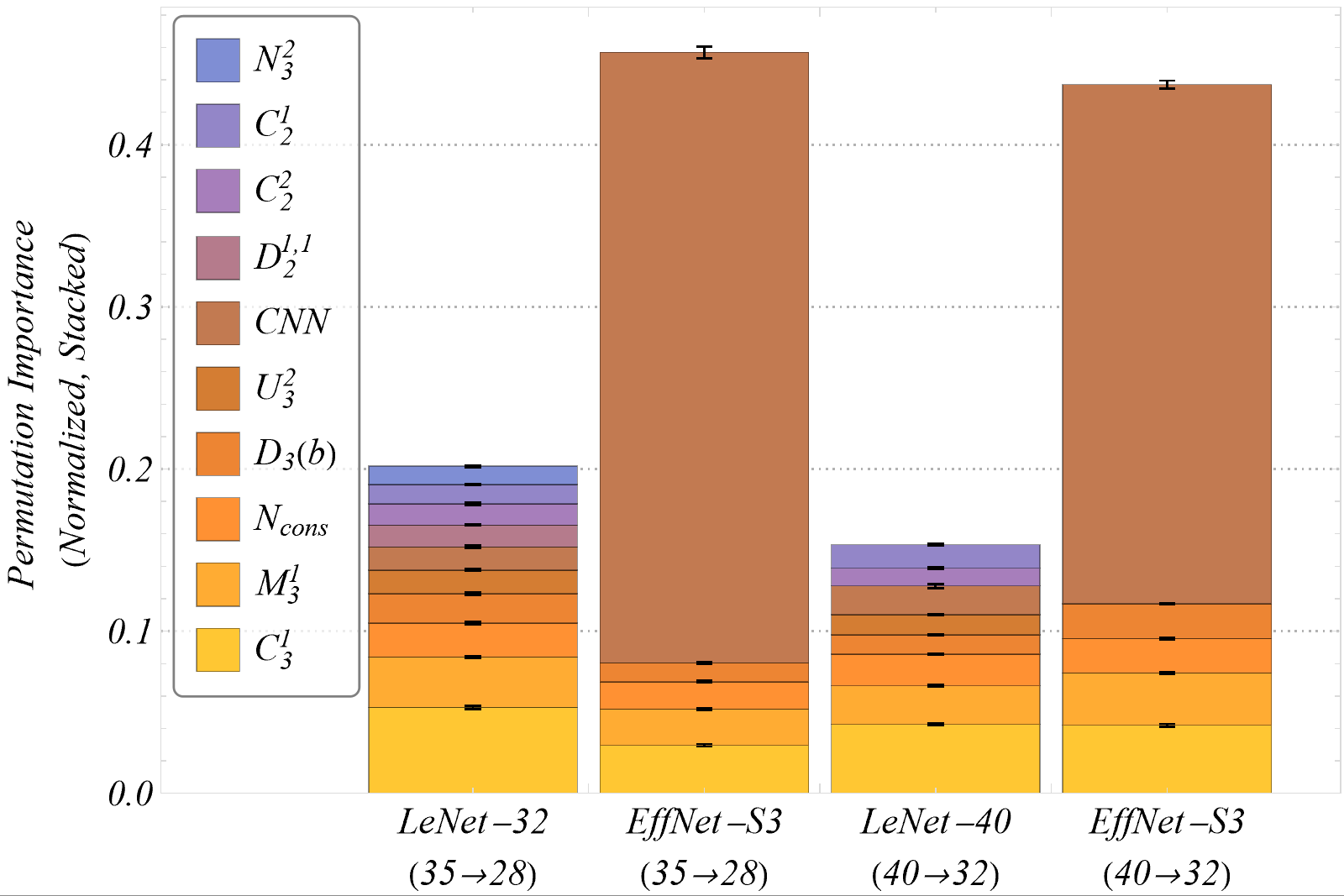}
        \caption{CNN vs Individual Global Features}
        \label{fig:importancePart}
    \end{subfigure}
    \caption{\small Comparison of relative importance of different CNN architectures for classification, in the presence of Global Features, using Normalized Permutation Importance Plots: fig. (\textit{a}) shows the CNN importance pitted against all the global features, taken together, fig. (\textit{b}) shows the most important features ($>1\%$), including the CNN output.}
    \label{fig:importance}
\end{figure*}

\subsection{Training}
\label{sec:top_training}
Although we load the validation and test datasets into RAM as a whole, due to our choice of confining ourselves to a single mid-range PC with $64 GB$ RAM, we use a type of data piping for the training dataset. We load $300,000$ training data at a time, which we will call a `\textit{round}' from now on. Each round is trained until the validation loss stops decreasing for 10 epochs. An image of the trained network is saved from each round and used in the next one. This is done for 12 rounds for $35\times 35$ ($40\times 40$) image representation of jets to ensure that the entire dataset is adequately represented. For most of the models used in this work, we repeat the process, starting with the last trained network, and use the same rounds again. This way, the rounds mimic the behavior of epochs, while storing the datasets in advance reduces computation time. We then evaluate the accuracy of each model on the test dataset, and the model with the highest accuracy is saved as the best version of that specific architecture. All further measurements are performed on this iteration of the model.

To ensure optimal computation time, we choose batch sizes depending on whether the memory dedicated to the graphics card can handle them without overflowing into the SWAP memory. For \textit{EffNet-S} versions $0$, $1$, and $2$, a batch size of $4096$ is used; for \textit{EffNet-S} versions $3$ and $4$, these numbers are respectively $3072$ and $2560$. For all other networks, it is fixed at $4096$. We use the \textit{ADAM} optimizer \cite{kingma2014adam} with default settings ($\beta_1 = 0.9$, $\beta_2 = 0.99$, and $\epsilon = 10^{-5}$, along with the default learning rate update schedule) to minimize the cross-entropy loss, where we wait for a fixed number of \textit{epochs} (generally 10) for the last minimum loss to be updated; otherwise, we stop the training.

The global features of the jets are standardized and utilized in the second part of the analysis, in conjunction with the images. These are concatenated with the output of the `\textit{Flatten Layer}' of \textit{LeNet} and, in the case of \textit{EffNet-S}, that of the `\textit{Aggregation Layer}', as can be seen from Fig. (\ref{fig:netdiag}). Pre-trained networks are faster to retrain and reach higher accuracies after modification—a fact that has been re-established in recent attempts at foundational models in jet applications
. Keeping that in mind, we use the first part of the best network of a particular architecture and add an MLP block to it in this part of the analysis. 

We retrain the best network of each architecture 9 additional times, with a representative subset of the dataset, to obtain the uncertainties arising from random weight initialization. 

\section{Results}
\label{sec:result}
\subsection{Performance}
Table (\ref{tab:pointres}) lists all single-number performance metrics for the networks explored in this analysis. The table is divided into three blocks: the first block contains all the information about the \textit{LeNet} architecture, the second one about \textit{EffNet-S}, and the third one lists some of the publicly available information about CNNs used previously as top-taggers for comparison. We have obtained results for \textit{LeNet}-type networks with three two different input resolutions: $35\times 35$, and $40\times 40$. Versions of \textit{EffNet-S} are of 5 types: $0$ to $4$. Images are generated in two ways: $35\times 35$ images cropped to $28\times 28$ and $40\times 40$ images cropped to $32\times 32$, as explained in sec. \ref{sec:data_images}. Each network is run on all images with cropped resolution (col. 3 of tab. \ref{tab:pointres}) less than or equal to its input dimension (col. 2 of tab. \ref{tab:pointres}). Networks with a specific architecture and input resolution optimized for maximum accuracy are used to create versions that can also process the global features. 

\textit{LeNets} with increasing input array-sizes, when applied only to images cropped to $28\times 28$, gain marginally in performance across all metrics. Incorporating global features clearly increases the discriminative power—especially background rejection—of the networks, and the effect is more prominent for higher image resolutions. This is expected, as \textit{LeNet} is a conventional CNN. Full convolutions with a large kernel size ($5$) extract more information from the images, albeit at a higher computational cost and a heavier network. The \textit{LeNet-40} is $2.9 (2.7)$ times the size of \textit{LeNet-40} without(with) global features.

The smallest \textit{EffNet-S}, while applied only to images, achieves accuracy similar to the largest \textit{LeNet}, while being $1/8$-th in size (the total number of parameters). Its performance is best when the input images are $35\xrightarrow{Crop} 28$, but it drops quite a lot for the larger images. Adding global features is beneficial here as well, but the effect is less prominent for larger networks. We have shown the best metrics in each column in bold-face, for both the cases with and without global features. 

As was mentioned earlier in sec. \ref{sec:machine}, the inference timings were calculated in a separate machine with a weaker GPU. While networks consistently take more time for larger image sizes, all of the networks dealt with in this work have almost half the inference time of that of ResNeXt-50. 

Fig. (\ref{fig:compnets}) summarizes the accuracies of table (\ref{tab:pointres}) in comparison to architecture complexity (number of net-parameters). Fig. (\ref{fig:compall}) shows the relative sizes of different versions of \textit{EffNet-S} compared to the versions of \textit{LeNet}s used in this analysis as well as the larger and standard \textit{EfficientNet}s (\textit{B0} to \textit{B7}), where the number of parameters is in log-scale. Even the largest \textit{EffNet-S} is smaller than the smallest \textit{LeNet} and all of these are an order of magnitude smaller than \textit{EffNet-B}s. In figures (\ref{fig:compbest}) and (\ref{fig:compdist}), accuracies of the nets are compared, with their model complexity in the $x$-axis. While fig. (\ref{fig:compbest}) plots the best accuracies for each type of net, fig. (\ref{fig:compdist}) shows the variation in it arising from random weight initialization. These employ 10 runs each, as mentioned at the end of the last section.

As far as accuracies are concerned and for only images as input, we see that the performance of \textit{LeNet} does not change much (and remains within randomization error), with increasing complexity, for all image resolutions. The \textit{EffNet-S}, on the other hand, performs worse for larger image sizes. For the larger variants, this loss in performance falls within the randomization error-bar, but overall, the performance not only worsens, but has larger variance for $40\xrightarrow{Crop} 32$ images. The possible reason is that the `MB-Conv' blocks with bottlenecks in \textit{EffNet-S} do not perform at the level of full convolutions when analyzing sparse pixel-level information, and networks with larger input arrays are needed for proper classification.

While the addition of global features improves network performances by a margin, the accuracies become more or less independent of the network complexity across architectures, especially when randomization errors are considered. One possibility is that the extra MLP block after the last Aggregation layer contributes far less than the rest of the network, but the more probable explanation would be that the impact of global features overshadow the CNN-performance. To analyze the importance of the global features relative to the base-CNN architectures, we have performed a permutation feature-importance analysis. Please see sec. \ref{sec:importance}.

Fig. (\ref{fig:roc}) shows the signal (top) efficiency vs. the background (QCD) rejection curves for those representative models for which at least one of the metrics is highest among others (those in bold-face in table \ref{tab:pointres}). This also shows how all \textit{EffNet-S} nets outperform the best \textit{LeNet} with just images as input (dashed curves), but with Global features put in, all of their performances are equivalent (solid, colored curves). We have also added the ROC curves of the best performing CNNs to date, namely \textit{ResNeXt-50} (black solid line) and \textit{DeepTop} (black dot-dashed line), for comparison. 

In the case of other CNNs, the entries for the `Image Resolution' column are calculated in the following way: we know these images were created $40\times 40$ (\textit{DeepTop}) and $64\times 64$ (for \textit{ResNext})  for $R = 0.8$. This corresponds to $60\times 60$ and $96\times 96$ for $R=1.2$, which is our case. These are all grayscale images with only one channel for $p_T$. All networks used in this analysis that utilize the global features outperform the classic \textit{DeepTop} CNN structure in terms of accuracy, if not in background rejection. \textit{ResNeXt-50} has a much better performance, but it is considerably heavier than the \textit{EffNet-S} (the best performing global feat. variant is $1/7$-th in size of \textit{ResNeXt-50}). These nets are faster as well. The longest inference time for \textit{EffNet-S} is half of that for \textit{ResNeXt-50}.

\subsection{Feature Importance}\label{sec:importance}
\textit{Permutation Feature Importance} is one of the many methods used to determine the impact any one feature has over the predictions of an ML model. Using a test dataset with known targets, one of the features is shuffled among the whole dataset and some performance metric is measured (we use Accuracy $S_i$). This process is repeated multiple times ($N$). As the shuffling jumbles the data, the performance of the model on the shuffled data is expected to be less than the actual performance. The difference between the base performance ($S_b$) and the average of $S_i$s is thus larger for an important feature, and is less for a less important one. This quantity is then Normalized over all ($P$) the features ($f_j$) to obtain the importance score.
\begin{equation}
    S(f_j) = \frac{S_b - \frac{1}{N}\sum_{i}^{N} S_i (f_j)}{\sum_{j}^{P}\left(S_b - \frac{1}{N}\sum_{i}^{N} S_i (f_j)\right)}
 \end{equation}
 Fig. \ref{fig:importance} shows the importance of the base CNN networks, relative to the global features---both together and individually. We see that for all \textit{LeNets} with global features present, most of the classification load is carried by the global features (fig. \ref{fig:importanceFull}). The CNN contribution is at most comparable to some high performing global features individually (fig. \ref{fig:importancePart}). On the other hand, though the global features together are still important in the \textit{EffNets}, the \textit{EffNet} contributes significantly in classification, outperforming even the best-performing global feature ($C_3^1$) by an order of magnitude.

\subsection{Proof of Concept: Scalability}\label{sec:pocS} 
To prove that the \textit{EffNet-S} series is expandable with compound scaling, we create one more network, \textit{EffNet-S5 (72)}, following the compound scaling explained in section \ref{sec:mlmodels}. At $515K$ trainable parameters, it is almost one-third the size of \textit{ResNeXt-50}. Table \ref{tab:effS5} summarizes the results we get when run on the $35\times 35 \to 28\times 28$ images. Compared to \textit{ResNeXt-50}, the accuracy is $0.1\%$ less, the background rejection at TPR $50\%$ is consistent within $1~\sigma$, but that at TPR $30\%$ is larger by more than $1~\sigma$. It clearly shows that \textit{EffNet-S} can easily outperform the SOTA CNN at less than half the model complexity. However, given our limited computational resources, we cannot scale the model further up. We leave that for others with access to better computational facility.
\begin{table}[!h]
    \centering
    \renewcommand{\arraystretch}{1.5}
    \begin{tabular}{lcc}
        \textbf{Measures} & \textbf{Best} & \textbf{Average} \\
        \hline
         Accuracy   & $93.49\%$         & $93.476(8)\%$\\
         AUC        & $98.32\%$         & $98.321(1)\%$ \\
         \multicolumn{3}{c}{Bkg. Rejection at TPR} \\
         $50\%$     &  & $294.4\pm 1.7$ \\
         $30\%$     &  & $1225\pm 18$
    \end{tabular}
    \caption{Performance of a scaled-up version of EffNet-S, \textit{S5(72)}.}
    \label{tab:effS5}
\end{table}

\section{Conclusion}
\label{sec:concl}
Jet tagging, a classification problem that aims to identify jets originating from different particles (top quarks, Higgs boson, gauge bosons, light quarks, or gluons), remains a challenging, albeit one of the most important tasks, especially in the context of a hadron collider like the LHC.
Machine learning has now replaced traditional use of hand-crafted features to become an integral part of jet physics studies. Depending on the network used, jets are treated as images, sequences, trees or graphs. Existing literature shows that transformer-based networks, which treat jets as particle clouds or graphs, perform the best. To achieve the highest possible accuracy, incorporating pairwise information is crucial. However, given the large number of pairwise interactions among the constituents of a jet, creating a fully connected graph becomes a computationally costly affair. CNNs, on the other hand, while being computationally low-cost, treat jets as images/arrays denoting the spatial arrangement of constituent particles in terms of energy, momentum, and mass; thus can utilize at least some local pairwise information. One can additionally incorporate global features of the jets to either directly to the images or to the network to enhance the jet identification accuracy. We have explored the possibility of combining these features and construct a computationally lightweight but efficient CNN-based network that can compete with other similar architectures.

With traditional CNNs, larger/deeper networks are needed to extract maximal information from the jet images, as opposed to typically lightweight CNNs like \textit{LeNet}. Scaled-up CNNs, such as \textit{ResNets} are computationally costly. Using recent advancements in the construction of CNNs, we have used lightweight versions of \textit{EfficientNet} to circumvent this issue. Although these were built to work on higher resolution images, one can follow a set of prescribed scaling rules to construct a network applicable for low resolution jet images as well. We have performed a comparative study with and without externally provided global features. Results for \textit{LeNet} are also provided as a benchmark for comparison.

We observe that adding global features to the models, along with the images, results in better accuracy. Although this the increase in performance is somewhat monotonic with increasing image resolution for fully convolutional models, such as \textit{LeNet}, the corresponding relationship for \textit{EffNet-S} is not so straightforward; increasing the input array size also improves the overall performance. The performance of these models becomes somewhat independent of their architecture in the presence of the global features. A permutation-feature-importance analysis reveals that some global features are so important (at least for the present dataset), that together they completely outperform conventional CNNs like \textit{LeNet}. \textit{EffNets}, on the other hand, contribute considerably to classification, making the overall performance better when they work in tandem with the global features. Further inquiry is required to analyze the scope of contribution of the global features in various jet-classifier architectures. 

This exploration leads to three observations: \textit{first}, there needs to be a systematic architecture search for versions of scalable models such as EfficientNet, tuned for jet classification; \textit{second}, adding global jet features, in addition to the constituent information, makes it easier for smaller models to perform better, but the exact way these features are fed-in needs further exploration; \textit{last but not least}, as jets can have multiple types of representation, e.g., image, constituent four-momenta, and global jet features, one can combine the strengths of different network architectures and build a hybrid `mixture of experts' ensemble of lightweight networks with the potential for superior performance, leading to better signal-to-background discrimination. Exploring these ideas is beyond the scope of this paper, but it is something that we plan to work on as a subsequent project.

\begin{acknowledgements}
RB, SM, and SKP would like to acknowledge support from ANRF, India (grant order no. CRG/2022/003208). SM would also like to acknowledge ANRF for grant order no.\\ CRG/2023/008570.
\end{acknowledgements}

\bibliographystyle{
utphys}
\bibliography{MLTOP}

@article{Yang:2025ktj,
    author = "Yang, Shuo and Wang, Yi-Hang and Zhao, Peng-Bo and Ma, Ji-Long",
    title = "{Searching for heavy vector-like B quark via pair production in fully hadronic channels at the CLIC}",
    eprint = "2504.15882",
    archivePrefix = "arXiv",
    primaryClass = "hep-ph",
    month = "4",
    year = "2025"
}

@article{Ghosh:2025gdq,
    author = "Ghosh, Anupam and Ghosh, Soumyadip and Mitra, Soureek and Samui, Tousik and Singh, Ritesh K.",
    title = "{Improving Sensitivity of Vector-like Top Partner Searches with Jet Substructure}",
    eprint = "2507.03199",
    archivePrefix = "arXiv",
    primaryClass = "hep-ph",
    month = "7",
    year = "2025"
}

@article{Chakraborty:2018khw,
    author = "Chakraborty, Sabyasachi and Mitra, Manimala and Shil, Sujay",
    title = "{Fat Jet Signature of a Heavy Neutrino at Lepton Collider}",
    eprint = "1810.08970",
    archivePrefix = "arXiv",
    primaryClass = "hep-ph",
    doi = "10.1103/PhysRevD.100.015012",
    journal = "Phys. Rev. D",
    volume = "100",
    number = "1",
    pages = "015012",
    year = "2019"
}

@article{Padhan:2022fak,
    author = "Padhan, Rojalin and Mitra, Manimala and Kulkarni, Suchita and Deppisch, Frank F.",
    title = "{Displaced fat-jets and tracks to probe boosted right-handed neutrinos in the $U(1)_{B-L}$ model}",
    eprint = "2203.06114",
    archivePrefix = "arXiv",
    primaryClass = "hep-ph",
    doi = "10.1140/epjc/s10052-022-10819-7",
    journal = "Eur. Phys. J. C",
    volume = "82",
    number = "10",
    pages = "858",
    year = "2022"
}

@article{Baruah:2024wrn,
    author = "Baruah, Rajneil and Choudhury, Arghya and Ghosh, Kirtiman and Mondal, Subhadeep and Sahu, Rameswar",
    title = "{Probing sub-TeV Higgsinos aided by a machine-learning-based top tagger in the context of trilinear R-parity violating SUSY}",
    eprint = "2412.11862",
    archivePrefix = "arXiv",
    primaryClass = "hep-ph",
    doi = "10.1103/PhysRevD.111.095004",
    journal = "Phys. Rev. D",
    volume = "111",
    number = "9",
    pages = "095004",
    year = "2025"
}

@article{Sahu:2025uop,
    author = "Sahu, Rameswar and Sahoo, Debabrata and Ghosh, Kirtiman",
    title = "{Advancing Higgsino Searches by Integrating ML for Boosted Object Tagging and Event Selection}",
    eprint = "2501.07491",
    archivePrefix = "arXiv",
    primaryClass = "hep-ph",
    month = "1",
    year = "2025"
}

@article{Sahoo:2025kdj,
    author = "Sahoo, Debabrata and Sahu, Rameswar and Ghosh, Kirtiman",
    title = "{A Comprehensive Search for Leptoquarks Decaying into Top-$\tau$ Final States at the Future LHC}",
    eprint = "2501.07543",
    archivePrefix = "arXiv",
    primaryClass = "hep-ph",
    month = "1",
    year = "2025"
}

@article{Bhaskar:2024snl,
    author = "Bhaskar, Arvind and Mitra, Manimala",
    title = "{Boosted top quark inspired leptoquark searches at the muon collider}",
    eprint = "2409.15992",
    archivePrefix = "arXiv",
    primaryClass = "hep-ph",
    doi = "10.1016/j.physletb.2025.139656",
    journal = "Phys. Lett. B",
    volume = "868",
    pages = "139656",
    year = "2025"
}

@article{Ghosh:2025gue,
    author = "Ghosh, Anupam and Konar, Partha and Samui, Tousik and Singh, Ritesh K.",
    title = "{Jet substructure probe on scalar leptoquark models via top polarization}",
    eprint = "2505.16328",
    archivePrefix = "arXiv",
    primaryClass = "hep-ph",
    reportNumber = "IMSc/2025/01",
    doi = "10.1007/JHEP07(2025)145",
    journal = "JHEP",
    volume = "07",
    pages = "145",
    year = "2025"
}

@article{Banerjee:2024zvg,
    author = "Banerjee, Avik and Bergeaas Kuutmann, Elin and Ellajosyula, Venugopal and Enberg, Rikard and Ferretti, Gabriele and Panizzi, Luca",
    title = "{Vector-like quarks: Status and new directions at the LHC}",
    eprint = "2406.09193",
    archivePrefix = "arXiv",
    primaryClass = "hep-ph",
    doi = "10.21468/SciPostPhysCore.7.4.079",
    journal = "SciPost Phys. Core",
    volume = "7",
    pages = "079",
    year = "2024"
}

@article{Choudhury:2021nib,
    author = "Choudhury, Debajyoti and Deka, Kuldeep and Kumar, Nilanjana",
    title = "{Looking for a vectorlike B quark at the LHC using jet substructure}",
    eprint = "2103.10655",
    archivePrefix = "arXiv",
    primaryClass = "hep-ph",
    doi = "10.1103/PhysRevD.104.035004",
    journal = "Phys. Rev. D",
    volume = "104",
    number = "3",
    pages = "035004",
    year = "2021"
}

@article{Chandak:2019iwj,
    author = "Chandak, Kushagra and Mandal, Tanumoy and Mitra, Subhadip",
    title = "{Hunting for scalar leptoquarks with boosted tops and light leptons}",
    eprint = "1907.11194",
    archivePrefix = "arXiv",
    primaryClass = "hep-ph",
    doi = "10.1103/PhysRevD.100.075019",
    journal = "Phys. Rev. D",
    volume = "100",
    number = "7",
    pages = "075019",
    year = "2019"
}

@article{Diaz:2017lit,
    author = "Diaz, Bastian and Schmaltz, Martin and Zhong, Yi-Ming",
    title = "{The leptoquark Hunter{\textquoteright}s guide: Pair production}",
    eprint = "1706.05033",
    archivePrefix = "arXiv",
    primaryClass = "hep-ph",
    doi = "10.1007/JHEP10(2017)097",
    journal = "JHEP",
    volume = "10",
    pages = "097",
    year = "2017"
}

@article{Abdullahi:2022jlv,
    author = "Abdullahi, Asli M. and others",
    title = "{The present and future status of heavy neutral leptons}",
    eprint = "2203.08039",
    archivePrefix = "arXiv",
    primaryClass = "hep-ph",
    reportNumber = "FERMILAB-CONF-22-184-T-V",
    doi = "10.1088/1361-6471/ac98f9",
    journal = "J. Phys. G",
    volume = "50",
    number = "2",
    pages = "020501",
    year = "2023"
}

@article{Bandyopadhyay:2022mej,
    author = "Bandyopadhyay, Priyotosh and Chun, Eung Jin and Sen, Chandrima",
    title = "{Boosted displaced decay of right-handed neutrinos at CMS, ATLAS and MATHUSLA}",
    eprint = "2205.12511",
    archivePrefix = "arXiv",
    primaryClass = "hep-ph",
    reportNumber = "IITH-PH-0002/22, KIAS-P22017",
    doi = "10.1007/JHEP02(2023)103",
    journal = "JHEP",
    volume = "02",
    pages = "103",
    year = "2023"
}

@article{Dreiner:2023bvs,
    author = {Dreiner, Herbi K. and Koay, Yong Sheng and K{\"o}hler, Dominik and Lozano, V{\'\i}ctor Mart{\'\i}n and Montejo Berlingen, Javier and Nangia, Saurabh and Strobbe, Nadja},
    title = "{The ABC of RPV: classification of R-parity violating signatures at the LHC for small couplings}",
    eprint = "2306.07317",
    archivePrefix = "arXiv",
    primaryClass = "hep-ph",
    reportNumber = "BONN-TH-2023-04, IFIC/23-21",
    doi = "10.1007/JHEP07(2023)215",
    journal = "JHEP",
    volume = "07",
    pages = "215",
    year = "2023"
}

@article{Dreiner:2025kfd,
    author = {Dreiner, Herbi K. and Hank, Michael and Koay, Yong Sheng and Sch{\"u}rmann, Martin and Sengupta, Rhitaja and Shah, Apoorva and Strobbe, Nadja and Thomson, Evelyn},
    title = "{The ABC of RPV. Part II. Classification of R-parity violating signatures from UDD couplings and their coverage at the LHC}",
    eprint = "2503.03830",
    archivePrefix = "arXiv",
    primaryClass = "hep-ph",
    reportNumber = "BONN-TH-2025-08",
    doi = "10.1007/JHEP06(2025)258",
    journal = "JHEP",
    volume = "06",
    pages = "258",
    year = "2025"
}

@article{Alekhin:2012py,
    author = "Alekhin, S. and Djouadi, A. and Moch, S.",
    title = "{The top quark and Higgs boson masses and the stability of the electroweak vacuum}",
    eprint = "1207.0980",
    archivePrefix = "arXiv",
    primaryClass = "hep-ph",
    reportNumber = "CERN-PH-TH-2012-186, DESY-12-113, LPT-ORSAY-12-64, LPN-12-073, SFB-CPP-12-41",
    doi = "10.1016/j.physletb.2012.08.024",
    journal = "Phys. Lett. B",
    volume = "716",
    pages = "214--219",
    year = "2012"
}

@article{Degrassi:2012ry,
    author = "Degrassi, Giuseppe and Di Vita, Stefano and Elias-Miro, Joan and Espinosa, Jose R. and Giudice, Gian F. and Isidori, Gino and Strumia, Alessandro",
    title = "{Higgs mass and vacuum stability in the Standard Model at NNLO}",
    eprint = "1205.6497",
    archivePrefix = "arXiv",
    primaryClass = "hep-ph",
    reportNumber = "CERN-PH-TH-2012-134, RM3-TH-12-9",
    doi = "10.1007/JHEP08(2012)098",
    journal = "JHEP",
    volume = "08",
    pages = "098",
    year = "2012"
}

@article{Elias-Miro:2011sqh,
    author = "Elias-Miro, Joan and Espinosa, Jose R. and Giudice, Gian F. and Isidori, Gino and Riotto, Antonio and Strumia, Alessandro",
    title = "{Higgs mass implications on the stability of the electroweak vacuum}",
    eprint = "1112.3022",
    archivePrefix = "arXiv",
    primaryClass = "hep-ph",
    doi = "10.1016/j.physletb.2012.02.013",
    journal = "Phys. Lett. B",
    volume = "709",
    pages = "222--228",
    year = "2012"
}

@article{Buttazzo:2013uya,
    author = "Buttazzo, Dario and Degrassi, Giuseppe and Giardino, Pier Paolo and Giudice, Gian F. and Sala, Filippo and Salvio, Alberto and Strumia, Alessandro",
    title = "{Investigating the near-criticality of the Higgs boson}",
    eprint = "1307.3536",
    archivePrefix = "arXiv",
    primaryClass = "hep-ph",
    reportNumber = "CERN-PH-TH-2013-166, FTUAM-13-20, IFT-UAM-CSIC-13-081, IFUP-TH",
    doi = "10.1007/JHEP12(2013)089",
    journal = "JHEP",
    volume = "12",
    pages = "089",
    year = "2013"
}

@article{Hiller:2024zjp,
    author = {Hiller, Gudrun and H{\"o}hne, Tim and Litim, Daniel F. and Steudtner, Tom},
    title = "{Vacuum stability in the Standard Model and beyond}",
    eprint = "2401.08811",
    archivePrefix = "arXiv",
    primaryClass = "hep-ph",
    doi = "10.1103/PhysRevD.110.115017",
    journal = "Phys. Rev. D",
    volume = "110",
    number = "11",
    pages = "115017",
    year = "2024"
}

@article{Espinosa:2015kwx,
    author = "Espinosa, Jose R.",
    editor = "Lista, Luca and Margaroli, Fabrizio and Tramontano, Francesco",
    title = "{Implications of the top (and Higgs) mass for vacuum stability}",
    eprint = "1512.01222",
    archivePrefix = "arXiv",
    primaryClass = "hep-ph",
    doi = "10.22323/1.257.0043",
    journal = "PoS",
    volume = "TOP2015",
    pages = "043",
    year = "2016"
}

@article{Mikuni:2024qsr,
    author = "Mikuni, Vinicius and Nachman, Benjamin",
    title = "{Solving key challenges in collider physics with foundation models}",
    eprint = "2404.16091",
    archivePrefix = "arXiv",
    primaryClass = "hep-ph",
    doi = "10.1103/PhysRevD.111.L051504",
    journal = "Phys. Rev. D",
    volume = "111",
    number = "5",
    pages = "L051504",
    year = "2025"
}

@article{Bhattacherjee:2022gjq,
    author = "Bhattacherjee, Biplob and Bose, Camellia and Chakraborty, Amit and Sengupta, Rhitaja",
    title = "{Boosted top tagging and its interpretation using Shapley values}",
    eprint = "2212.11606",
    archivePrefix = "arXiv",
    primaryClass = "hep-ph",
    doi = "10.1140/epjp/s13360-024-05910-9",
    journal = "Eur. Phys. J. Plus",
    volume = "139",
    number = "12",
    pages = "1131",
    year = "2024"
}

@article{Sahu:2024fzi,
    author = "Sahu, Rameswar and Ashanujjaman, Saiyad and Ghosh, Kirtiman",
    title = "{Unveiling the secrets of new physics through top quark tagging}",
    eprint = "2409.12085",
    archivePrefix = "arXiv",
    primaryClass = "hep-ph",
    doi = "10.1140/epjs/s11734-024-01257-5",
    journal = "Eur. Phys. J. ST",
    volume = "233",
    number = "15-16",
    pages = "2465--2495",
    year = "2024"
}

@article{Baldi:2016fql,
    author = "Baldi, Pierre and Bauer, Kevin and Eng, Clara and Sadowski, Peter and Whiteson, Daniel",
    title = "{Jet Substructure Classification in High-Energy Physics with Deep Neural Networks}",
    eprint = "1603.09349",
    archivePrefix = "arXiv",
    primaryClass = "hep-ex",
    doi = "10.1103/PhysRevD.93.094034",
    journal = "Phys. Rev. D",
    volume = "93",
    number = "9",
    pages = "094034",
    year = "2016"
}

@article{Kagan:2020yrm,
    author = "Kagan, Michael",
    title = "{Image-Based Jet Analysis}",
    eprint = "2012.09719",
    archivePrefix = "arXiv",
    primaryClass = "physics.data-an",
    month = "12",
    year = "2020"
}

@article{Kasieczka:2017nvn,
    author = "Kasieczka, Gregor and Plehn, Tilman and Russell, Michael and Schell, Torben",
    title = "{Deep-learning Top Taggers or The End of QCD?}",
    eprint = "1701.08784",
    archivePrefix = "arXiv",
    primaryClass = "hep-ph",
    reportNumber = "MCNET-17-07",
    doi = "10.1007/JHEP05(2017)006",
    journal = "JHEP",
    volume = "05",
    pages = "006",
    year = "2017"
}

@article{Macaluso:2018tck,
    author = "Macaluso, Sebastian and Shih, David",
    title = "{Pulling Out All the Tops with Computer Vision and Deep Learning}",
    eprint = "1803.00107",
    archivePrefix = "arXiv",
    primaryClass = "hep-ph",
    doi = "10.1007/JHEP10(2018)121",
    journal = "JHEP",
    volume = "10",
    pages = "121",
    year = "2018"
}

@article{Kasieczka:2019dbj,
    author = "Butter, Anja and others",
    editor = "Kasieczka, Gregor and Plehn, Tilman",
    title = "{The Machine Learning landscape of top taggers}",
    eprint = "1902.09914",
    archivePrefix = "arXiv",
    primaryClass = "hep-ph",
    doi = "10.21468/SciPostPhys.7.1.014",
    journal = "SciPost Phys.",
    volume = "7",
    pages = "014",
    year = "2019"
}

@article{Diefenbacher:2019ezd,
    author = "Diefenbacher, Sascha and Frost, Hermann and Kasieczka, Gregor and Plehn, Tilman and Thompson, Jennifer M.",
    title = "{CapsNets Continuing the Convolutional Quest}",
    eprint = "1906.11265",
    archivePrefix = "arXiv",
    primaryClass = "hep-ph",
    doi = "10.21468/SciPostPhys.8.2.023",
    journal = "SciPost Phys.",
    volume = "8",
    pages = "023",
    year = "2020"
}

@article{Bollweg:2019skg,
    author = "Bollweg, Sven and Hau{\ss}mann, Manuel and Kasieczka, Gregor and Luchmann, Michel and Plehn, Tilman and Thompson, Jennifer",
    title = "{Deep-Learning Jets with Uncertainties and More}",
    eprint = "1904.10004",
    archivePrefix = "arXiv",
    primaryClass = "hep-ph",
    doi = "10.21468/SciPostPhys.8.1.006",
    journal = "SciPost Phys.",
    volume = "8",
    number = "1",
    pages = "006",
    year = "2020"
}

@article{Butter:2017cot,
    author = "Butter, Anja and Kasieczka, Gregor and Plehn, Tilman and Russell, Michael",
    title = "{Deep-learned Top Tagging with a Lorentz Layer}",
    eprint = "1707.08966",
    archivePrefix = "arXiv",
    primaryClass = "hep-ph",
    doi = "10.21468/SciPostPhys.5.3.028",
    journal = "SciPost Phys.",
    volume = "5",
    number = "3",
    pages = "028",
    year = "2018"
}

@article{Shlomi:2020gdn,
    author = "Shlomi, Jonathan and Battaglia, Peter and Vlimant, Jean-Roch",
    title = "{Graph Neural Networks in Particle Physics}",
    eprint = "2007.13681",
    archivePrefix = "arXiv",
    primaryClass = "hep-ex",
    doi = "10.1088/2632-2153/abbf9a",
    month = "7",
    year = "2020"
}

@article{Duarte:2020ngm,
    author = "Duarte, Javier and Vlimant, Jean-Roch",
    title = "{Graph Neural Networks for Particle Tracking and Reconstruction}",
    eprint = "2012.01249",
    archivePrefix = "arXiv",
    primaryClass = "hep-ph",
    doi = "10.1142/9789811234033_0012",
    month = "12",
    year = "2020"
}

@inproceedings{Thais:2022iok,
    author = "Thais, Savannah and Calafiura, Paolo and Chachamis, Grigorios and DeZoort, Gage and Duarte, Javier and Ganguly, Sanmay and Kagan, Michael and Murnane, Daniel and Neubauer, Mark S. and Terao, Kazuhiro",
    title = "{Graph Neural Networks in Particle Physics: Implementations, Innovations, and Challenges}",
    booktitle = "{Snowmass 2021}",
    eprint = "2203.12852",
    archivePrefix = "arXiv",
    primaryClass = "hep-ex",
    month = "3",
    year = "2022"
}

@article{Komiske:2018cqr,
    author = "Komiske, Patrick T. and Metodiev, Eric M. and Thaler, Jesse",
    title = "{Energy Flow Networks: Deep Sets for Particle Jets}",
    eprint = "1810.05165",
    archivePrefix = "arXiv",
    primaryClass = "hep-ph",
    reportNumber = "MIT-CTP 5064",
    doi = "10.1007/JHEP01(2019)121",
    journal = "JHEP",
    volume = "01",
    pages = "121",
    year = "2019"
}

@article{Qu:2019gqs,
    author = "Qu, Huilin and Gouskos, Loukas",
    title = "{ParticleNet: Jet Tagging via Particle Clouds}",
    eprint = "1902.08570",
    archivePrefix = "arXiv",
    primaryClass = "hep-ph",
    doi = "10.1103/PhysRevD.101.056019",
    journal = "Phys. Rev. D",
    volume = "101",
    number = "5",
    pages = "056019",
    year = "2020"
}

@article{Bogatskiy:2020tje,
    author = "Bogatskiy, Alexander and Anderson, Brandon and Offermann, Jan T. and Roussi, Marwah and Miller, David W. and Kondor, Risi",
    title = "{Lorentz Group Equivariant Neural Network for Particle Physics}",
    eprint = "2006.04780",
    archivePrefix = "arXiv",
    primaryClass = "hep-ph",
    month = "6",
    year = "2020"
}

@article{Gong:2022lye,
    author = "Gong, Shiqi and Meng, Qi and Zhang, Jue and Qu, Huilin and Li, Congqiao and Qian, Sitian and Du, Weitao and Ma, Zhi-Ming and Liu, Tie-Yan",
    title = "{An efficient Lorentz equivariant graph neural network for jet tagging}",
    eprint = "2201.08187",
    archivePrefix = "arXiv",
    primaryClass = "hep-ph",
    doi = "10.1007/JHEP07(2022)030",
    journal = "JHEP",
    volume = "07",
    pages = "030",
    year = "2022"
}

@article{Bogatskiy:2023nnw,
    author = "Bogatskiy, Alexander and Hoffman, Timothy and Miller, David W. and Offermann, Jan T. and Liu, Xiaoyang",
    title = "{Explainable equivariant neural networks for particle physics: PELICAN}",
    eprint = "2307.16506",
    archivePrefix = "arXiv",
    primaryClass = "hep-ph",
    doi = "10.1007/JHEP03(2024)113",
    journal = "JHEP",
    volume = "03",
    pages = "113",
    year = "2024"
}

@article{Bogatskiy:2022czk,
    author = "Bogatskiy, Alexander and Hoffman, Timothy and Miller, David W. and Offermann, Jan T.",
    title = "{PELICAN: Permutation Equivariant and Lorentz Invariant or Covariant Aggregator Network for Particle Physics}",
    eprint = "2211.00454",
    archivePrefix = "arXiv",
    primaryClass = "hep-ph",
    month = "11",
    year = "2022"
}

@inproceedings{Bogatskiy:2023fug,
    author = "Bogatskiy, Alexander and Hoffman, Timothy and Offermann, Jan T.",
    title = "{19 Parameters Is All You Need: Tiny Neural Networks for Particle Physics}",
    booktitle = "{37th Conference on Neural Information Processing Systems}",
    eprint = "2310.16121",
    archivePrefix = "arXiv",
    primaryClass = "hep-ph",
    month = "10",
    year = "2023"
}

@article{DBLP:journals/corr/VaswaniSPUJGKP17,
  author       = {Ashish Vaswani and
                  Noam Shazeer and
                  Niki Parmar and
                  Jakob Uszkoreit and
                  Llion Jones and
                  Aidan N. Gomez and
                  Lukasz Kaiser and
                  Illia Polosukhin},
  title        = {Attention Is All You Need},
  journal      = {CoRR},
  volume       = {abs/1706.03762},
  year         = {2017},
  url          = {http://arxiv.org/abs/1706.03762},
  eprinttype    = {arXiv},
  eprint       = {1706.03762},
  bibsource    = {dblp computer science bibliography, https://dblp.org}
}

@ARTICLE{JoshiTrans2GNN,
       author = {{Joshi}, Chaitanya K.},
        title = "{Transformers are Graph Neural Networks}",
      journal = {arXiv e-prints},
         year = 2025,
        month = jun,
          eid = {arXiv:2506.22084},
        pages = {arXiv:2506.22084},
          doi = {10.48550/arXiv.2506.22084},
archivePrefix = {arXiv},
       eprint = {2506.22084},
 primaryClass = {stat.ML},
       adsurl = {https://ui.adsabs.harvard.edu/abs/2025arXiv250622084J}
}

@article{Qu:2022mxj,
    author = "Qu, Huilin and Li, Congqiao and Qian, Sitian",
    title = "{Particle Transformer for Jet Tagging}",
    eprint = "2202.03772",
    archivePrefix = "arXiv",
    primaryClass = "hep-ph",
    month = "2",
    year = "2022"
}

@article{Bhimji:2025isp,
    author = "Bhimji, Wahid and Harris, Chris and Mikuni, Vinicius and Nachman, Benjamin",
    title = "{OmniLearned: A Foundation Model Framework for All Tasks Involving Jet Physics}",
    eprint = "2510.24066",
    archivePrefix = "arXiv",
    primaryClass = "hep-ph",
    month = "10",
    year = "2025"
}

@article{Mikuni:2025tar,
    author = "Mikuni, Vinicius and Nachman, Benjamin",
    title = "{Method to simultaneously facilitate all jet physics tasks}",
    eprint = "2502.14652",
    archivePrefix = "arXiv",
    primaryClass = "hep-ph",
    doi = "10.1103/PhysRevD.111.054015",
    journal = "Phys. Rev. D",
    volume = "111",
    number = "5",
    pages = "054015",
    year = "2025"
}

@article{Mikuni:2021pou,
    author = "Mikuni, Vinicius and Canelli, Florencia",
    title = "{Point cloud transformers applied to collider physics}",
    eprint = "2102.05073",
    archivePrefix = "arXiv",
    primaryClass = "physics.data-an",
    doi = "10.1088/2632-2153/ac07f6",
    journal = "Mach. Learn. Sci. Tech.",
    volume = "2",
    number = "3",
    pages = "035027",
    year = "2021"
}

@inproceedings{Spinner:2024hjm,
    author = "Spinner, Jonas and Bres{\'o}, Victor and de Haan, Pim and Plehn, Tilman and Thaler, Jesse and Brehmer, Johann",
    title = "{Lorentz-Equivariant Geometric Algebra Transformers for High-Energy Physics}",
    booktitle = "{38th conference on Neural Information Processing Systems}",
    eprint = "2405.14806",
    archivePrefix = "arXiv",
    primaryClass = "physics.data-an",
    reportNumber = "MIT-CTP/5723",
    month = "10",
    year = "2024"
}

@article{Brehmer:2024yqw,
    author = "Brehmer, Johann and Bres{\'o}, V{\'\i}ctor and de Haan, Pim and Plehn, Tilman and Qu, Huilin and Spinner, Jonas and Thaler, Jesse",
    title = "{A Lorentz-equivariant transformer for all of the LHC}",
    eprint = "2411.00446",
    archivePrefix = "arXiv",
    primaryClass = "hep-ph",
    reportNumber = "MIT-CTP/5802",
    doi = "10.21468/SciPostPhys.19.4.108",
    journal = "SciPost Phys.",
    volume = "19",
    number = "4",
    pages = "108",
    year = "2025"
}

@article{Wu:2024thh,
    author = "Wu, Yifan and Wang, Kun and Li, Congqiao and Qu, Huilin and Zhu, Jingya",
    title = "{Jet tagging with more-interaction particle transformer*}",
    eprint = "2407.08682",
    archivePrefix = "arXiv",
    primaryClass = "hep-ph",
    doi = "10.1088/1674-1137/ad7f3d",
    journal = "Chin. Phys. C",
    volume = "49",
    number = "1",
    pages = "013110",
    year = "2025"
}

@article{Thaler:2010tr,
    author = "Thaler, Jesse and Van Tilburg, Ken",
    title = "{Identifying Boosted Objects with N-subjettiness}",
    eprint = "1011.2268",
    archivePrefix = "arXiv",
    primaryClass = "hep-ph",
    reportNumber = "MIT-CTP-4191",
    doi = "10.1007/JHEP03(2011)015",
    journal = "JHEP",
    volume = "03",
    pages = "015",
    year = "2011"
}

@article{Thaler:2011gf,
    author = "Thaler, Jesse and Van Tilburg, Ken",
    title = "{Maximizing Boosted Top Identification by Minimizing N-subjettiness}",
    eprint = "1108.2701",
    archivePrefix = "arXiv",
    primaryClass = "hep-ph",
    reportNumber = "MIT-CTP-4287",
    doi = "10.1007/JHEP02(2012)093",
    journal = "JHEP",
    volume = "02",
    pages = "093",
    year = "2012"
}

@article{Stewart:2010tn,
    author = "Stewart, Iain W. and Tackmann, Frank J. and Waalewijn, Wouter J.",
    title = "{N-Jettiness: An Inclusive Event Shape to Veto Jets}",
    eprint = "1004.2489",
    archivePrefix = "arXiv",
    primaryClass = "hep-ph",
    reportNumber = "MIT-CTP-4139",
    doi = "10.1103/PhysRevLett.105.092002",
    journal = "Phys. Rev. Lett.",
    volume = "105",
    pages = "092002",
    year = "2010"
}

@article{Larkoski:2013eya,
    author = "Larkoski, Andrew J. and Salam, Gavin P. and Thaler, Jesse",
    title = "{Energy Correlation Functions for Jet Substructure}",
    eprint = "1305.0007",
    archivePrefix = "arXiv",
    primaryClass = "hep-ph",
    reportNumber = "MIT-CTP-4446, CERN-PH-TH-2013-066, LPN13-026",
    doi = "10.1007/JHEP06(2013)108",
    journal = "JHEP",
    volume = "06",
    pages = "108",
    year = "2013"
}

@article{Moult:2016cvt,
    author = "Moult, Ian and Necib, Lina and Thaler, Jesse",
    title = "{New Angles on Energy Correlation Functions}",
    eprint = "1609.07483",
    archivePrefix = "arXiv",
    primaryClass = "hep-ph",
    reportNumber = "MIT-CTP-4825",
    doi = "10.1007/JHEP12(2016)153",
    journal = "JHEP",
    volume = "12",
    pages = "153",
    year = "2016"
}

@article{Komiske:2017aww,
    author = "Komiske, Patrick T. and Metodiev, Eric M. and Thaler, Jesse",
    title = "{Energy flow polynomials: A complete linear basis for jet substructure}",
    eprint = "1712.07124",
    archivePrefix = "arXiv",
    primaryClass = "hep-ph",
    reportNumber = "MIT-CTP-4965",
    doi = "10.1007/JHEP04(2018)013",
    journal = "JHEP",
    volume = "04",
    pages = "013",
    year = "2018"
}

@article{Abdesselam:2010pt,
    author = "Abdesselam, A. and others",
    title = "{Boosted Objects: A Probe of Beyond the Standard Model Physics}",
    eprint = "1012.5412",
    archivePrefix = "arXiv",
    primaryClass = "hep-ph",
    reportNumber = "SLAC-PUB-15081, FERMILAB-PUB-10-617-CMS",
    doi = "10.1140/epjc/s10052-011-1661-y",
    journal = "Eur. Phys. J. C",
    volume = "71",
    pages = "1661",
    year = "2011"
}

@article{Kogler:2018hem,
    author = "Kogler, Roman and others",
    title = "{Jet Substructure at the Large Hadron Collider: Experimental Review}",
    eprint = "1803.06991",
    archivePrefix = "arXiv",
    primaryClass = "hep-ex",
    reportNumber = "FERMILAB-PUB-18-123-PPD",
    doi = "10.1103/RevModPhys.91.045003",
    journal = "Rev. Mod. Phys.",
    volume = "91",
    number = "4",
    pages = "045003",
    year = "2019"
}

@article{Altheimer:2012mn,
    author = "Altheimer, A. and others",
    title = "{Jet Substructure at the Tevatron and LHC: New Results, New Tools, New Benchmarks}",
    eprint = "1201.0008",
    archivePrefix = "arXiv",
    primaryClass = "hep-ph",
    reportNumber = "SLAC-R-990, FERMILAB-PUB-12-897-T",
    doi = "10.1088/0954-3899/39/6/063001",
    journal = "J. Phys. G",
    volume = "39",
    pages = "063001",
    year = "2012"
}

@article{Altheimer:2013yza,
    author = "Altheimer, A. and others",
    title = "{Boosted Objects and Jet Substructure at the LHC. Report of BOOST2012, held at IFIC Valencia, 23rd-27th of July 2012}",
    eprint = "1311.2708",
    archivePrefix = "arXiv",
    primaryClass = "hep-ex",
    reportNumber = "FERMILAB-PUB-13-669",
    doi = "10.1140/epjc/s10052-014-2792-8",
    journal = "Eur. Phys. J. C",
    volume = "74",
    number = "3",
    pages = "2792",
    year = "2014"
}

@article{Adams:2015hiv,
    author = "Adams, D. and others",
    title = "{Towards an Understanding of the Correlations in Jet Substructure}",
    eprint = "1504.00679",
    archivePrefix = "arXiv",
    primaryClass = "hep-ph",
    reportNumber = "FERMILAB-PUB-15-670-CMS, SLAC-PUB-16703",
    doi = "10.1140/epjc/s10052-015-3587-2",
    journal = "Eur. Phys. J. C",
    volume = "75",
    number = "9",
    pages = "409",
    year = "2015"
}

@book{Marzani:2019hun,
    author = "Marzani, Simone and Soyez, Gregory and Spannowsky, Michael",
    title = "{Looking inside jets: an introduction to jet substructure and boosted-object phenomenology}",
    eprint = "1901.10342",
    archivePrefix = "arXiv",
    primaryClass = "hep-ph",
    doi = "10.1007/978-3-030-15709-8",
    publisher = "Springer",
    volume = "958",
    year = "2019"
}

@article{Larkoski:2017jix,
    author = "Larkoski, Andrew J. and Moult, Ian and Nachman, Benjamin",
    title = "{Jet Substructure at the Large Hadron Collider: A Review of Recent Advances in Theory and Machine Learning}",
    eprint = "1709.04464",
    archivePrefix = "arXiv",
    primaryClass = "hep-ph",
    doi = "10.1016/j.physrep.2019.11.001",
    journal = "Phys. Rept.",
    volume = "841",
    pages = "1--63",
    year = "2020"
}

@article{deFavereau:2013fsa,
    author = "de Favereau, J. and Delaere, C. and Demin, P. and Giammanco, A. and Lema\^\i{}tre, V. and Mertens, A. and Selvaggi, M.",
    collaboration = "DELPHES 3",
    title = "{DELPHES 3, A modular framework for fast simulation of a generic collider experiment}",
    eprint = "1307.6346",
    archivePrefix = "arXiv",
    primaryClass = "hep-ex",
    doi = "10.1007/JHEP02(2014)057",
    journal = "JHEP",
    volume = "02",
    pages = "057",
    year = "2014"
}

@article{Cacciari:2008gp,
    author = "Cacciari, Matteo and Salam, Gavin P. and Soyez, Gregory",
    title = "{The anti-$k_t$ jet clustering algorithm}",
    eprint = "0802.1189",
    archivePrefix = "arXiv",
    primaryClass = "hep-ph",
    reportNumber = "LPTHE-07-03",
    doi = "10.1088/1126-6708/2008/04/063",
    journal = "JHEP",
    volume = "04",
    pages = "063",
    year = "2008"
}

@article{Cacciari:2011ma,
    author = "Cacciari, Matteo and Salam, Gavin P. and Soyez, Gregory",
    title = "{FastJet User Manual}",
    eprint = "1111.6097",
    archivePrefix = "arXiv",
    primaryClass = "hep-ph",
    reportNumber = "CERN-PH-TH-2011-297",
    doi = "10.1140/epjc/s10052-012-1896-2",
    journal = "Eur. Phys. J. C",
    volume = "72",
    pages = "1896",
    year = "2012"
}

@article{Sjostrand:2014zea,
    author = {Sj{\"o}strand, Torbj{\"o}rn and Ask, Stefan and Christiansen, Jesper R. and Corke, Richard and Desai, Nishita and Ilten, Philip and Mrenna, Stephen and Prestel, Stefan and Rasmussen, Christine O. and Skands, Peter Z.},
    title = "{An introduction to PYTHIA 8.2}",
    eprint = "1410.3012",
    archivePrefix = "arXiv",
    primaryClass = "hep-ph",
    reportNumber = "LU-TP-14-36, MCNET-14-22, CERN-PH-TH-2014-190, FERMILAB-PUB-14-316-CD, DESY-14-178, SLAC-PUB-16122",
    doi = "10.1016/j.cpc.2015.01.024",
    journal = "Comput. Phys. Commun.",
    volume = "191",
    pages = "159--177",
    year = "2015"
}

@ARTICLE{xyzcolorCEI:1931,
       author = {{Smith}, T. and {Guild}, J.},
        title = "{The C.I.E. colorimetric standards and their use}",
      journal = {Transactions of the Optical Society},
         year = 1931,
        month = jan,
       volume = {33},
       number = {3},
        pages = {73-134},
          doi = {10.1088/1475-4878/33/3/301},
       adsurl = {https://ui.adsabs.harvard.edu/abs/1931TrOS...33...73S}
}

@misc{effnet:Wolfram,
  author = {M. Tan, Q. Le},
  title = "{EfficientNet Trained on ImageNet}",
  year = {2020},
  url = {https://resources.wolframcloud.com/NeuralNetRepository/resources/EfficientNet-Trained-on-ImageNet/},
  note = {[Accessed: 30-September-2025]}
}

@article{kingma2014adam,
  title={Adam: A method for stochastic optimization},
  author={Kingma, Diederik P},
  journal={arXiv preprint arXiv:1412.6980},
  year={2014}
}

@ARTICLE{lenetPaper,
  author={Lecun, Y. and Bottou, L. and Bengio, Y. and Haffner, P.},
  journal={Proceedings of the IEEE}, 
  title={Gradient-based learning applied to document recognition}, 
  year={1998},
  volume={86},
  number={11},
  pages={2278-2324},
  doi={10.1109/5.726791}}

@inproceedings{AlexNetPaper,
 author = {Krizhevsky, Alex and Sutskever, Ilya and Hinton, Geoffrey E},
 booktitle = {Advances in Neural Information Processing Systems},
 editor = {F. Pereira and C.J. Burges and L. Bottou and K.Q. Weinberger},
 pages = {},
 publisher = {Curran Associates, Inc.},
 title = {ImageNet Classification with Deep Convolutional Neural Networks},
 url = {https://proceedings.neurips.cc/paper_files/paper/2012/file/c399862d3b9d6b76c8436e924a68c45b-Paper.pdf},
 volume = {25},
 year = {2012}
}

@INPROCEEDINGS{InceptionV1,
  author={Szegedy, Christian and Wei Liu and Yangqing Jia and Sermanet, Pierre and Reed, Scott and Anguelov, Dragomir and Erhan, Dumitru and Vanhoucke, Vincent and Rabinovich, Andrew},
  booktitle={2015 IEEE Conference on Computer Vision and Pattern Recognition (CVPR)}, 
  title={Going deeper with convolutions}, 
  year={2015},
  volume={},
  number={},
  pages={1-9},
  doi={10.1109/CVPR.2015.7298594}}

@INPROCEEDINGS{InceptionV2V3,
  author={Szegedy, Christian and Vanhoucke, Vincent and Ioffe, Sergey and Shlens, Jon and Wojna, Zbigniew},
  booktitle={2016 IEEE Conference on Computer Vision and Pattern Recognition (CVPR)}, 
  title={Rethinking the Inception Architecture for Computer Vision}, 
  year={2016},
  volume={},
  number={},
  pages={2818-2826},
  doi={10.1109/CVPR.2016.308}}

@article{InceptionV4,
    title={Inception-v4, Inception-ResNet and the Impact of Residual Connections on Learning}, 
    volume={31}, 
    url={https://ojs.aaai.org/index.php/AAAI/article/view/11231}, 
    DOI={10.1609/aaai.v31i1.11231}, 
    number={1}, 
    journal={Proceedings of the AAAI Conference on Artificial Intelligence}, 
    author={Szegedy, Christian and Ioffe, Sergey and Vanhoucke, Vincent and Alemi, Alexander}, 
    year={2017}, 
    month={Feb.}}

@INPROCEEDINGS{ResNet1,
  author={He, Kaiming and Zhang, Xiangyu and Ren, Shaoqing and Sun, Jian},
  booktitle={2016 IEEE Conference on Computer Vision and Pattern Recognition (CVPR)}, 
  title={Deep Residual Learning for Image Recognition}, 
  year={2016},
  volume={},
  number={},
  pages={770-778},
  doi={10.1109/CVPR.2016.90}}

@article{wideResNet,
  title={Wide residual networks},
  author={Zagoruyko, Sergey and Komodakis, Nikos},
  journal={arXiv preprint arXiv:1605.07146},
  year={2016}
}

@INPROCEEDINGS{DenseNet,
  author={Huang, Gao and Liu, Zhuang and Van Der Maaten, Laurens and Weinberger, Kilian Q.},
  booktitle={2017 IEEE Conference on Computer Vision and Pattern Recognition (CVPR)}, 
  title={Densely Connected Convolutional Networks}, 
  year={2017},
  volume={},
  number={},
  pages={2261-2269},
  doi={10.1109/CVPR.2017.243}}

@INPROCEEDINGS{ResNext,
  author={Xie, Saining and Girshick, Ross and Dollár, Piotr and Tu, Zhuowen and He, Kaiming},
  booktitle={2017 IEEE Conference on Computer Vision and Pattern Recognition (CVPR)}, 
  title={Aggregated Residual Transformations for Deep Neural Networks}, 
  year={2017},
  volume={},
  number={},
  pages={5987-5995},
  doi={10.1109/CVPR.2017.634}}

@article{MobileNetV1,
  title={Mobilenets: Efficient convolutional neural networks for mobile vision applications},
  author={Howard, Andrew G and Zhu, Menglong and Chen, Bo and Kalenichenko, Dmitry and Wang, Weijun and Weyand, Tobias and Andreetto, Marco and Adam, Hartwig},
  journal={arXiv preprint arXiv:1704.04861},
  year={2017}
}

@inproceedings{MobileNetV2,
  title={Mobilenetv2: Inverted residuals and linear bottlenecks},
  author={Sandler, Mark and Howard, Andrew and Zhu, Menglong and Zhmoginov, Andrey and Chen, Liang-Chieh},
  booktitle={Proceedings of the IEEE conference on computer vision and pattern recognition},
  pages={4510--4520},
  year={2018}
}

@inproceedings{MobileNetV3,
  title={Searching for mobilenetv3},
  author={Howard, Andrew and Sandler, Mark and Chu, Grace and Chen, Liang-Chieh and Chen, Bo and Tan, Mingxing and Wang, Weijun and Zhu, Yukun and Pang, Ruoming and Vasudevan, Vijay and others},
  booktitle={Proceedings of the IEEE/CVF international conference on computer vision},
  pages={1314--1324},
  year={2019}
}

@inproceedings{MobileNetV4,
  title={MobileNetV4: Universal models for the mobile ecosystem},
  author={Qin, Danfeng and Leichner, Chas and Delakis, Manolis and Fornoni, Marco and Luo, Shixin and Yang, Fan and Wang, Weijun and Banbury, Colby and Ye, Chengxi and Akin, Berkin and others},
  booktitle={European Conference on Computer Vision},
  pages={78--96},
  year={2024},
  organization={Springer}
}

@inproceedings{MNASnet,
  title={Mnasnet: Platform-aware neural architecture search for mobile},
  author={Tan, Mingxing and Chen, Bo and Pang, Ruoming and Vasudevan, Vijay and Sandler, Mark and Howard, Andrew and Le, Quoc V},
  booktitle={Proceedings of the IEEE/CVF conference on computer vision and pattern recognition},
  pages={2820--2828},
  year={2019}
}

@inproceedings{ImageNet2019,
    title={Towards Fairer Datasets: Filtering and Balancing the Distribution of the People Subtree in the ImageNet Hierarchy},
    author={Kaiyu Yang and Klint Qinami and Li Fei-Fei and Jia Deng and Olga Russakovsky},
    booktitle={Conference on Fairness, Accountability, and Transparency},
    doi={10.1145/3351095.3375709},
    year={2020}
}

@inproceedings{ImageNet2021,
 title={A Study of Face Obfuscation in ImageNet},
 author={Yang, Kaiyu and Yau, Jacqueline and Fei-Fei, Li and Deng, Jia and Russakovsky, Olga},
 booktitle={International Conference on Machine Learning (ICML)},
 year={2022}
}

@inproceedings{long2015fully,
  title={Fully convolutional networks for semantic segmentation},
  author={Long, Jonathan and Shelhamer, Evan and Darrell, Trevor},
  booktitle={Proceedings of the IEEE conference on computer vision and pattern recognition},
  pages={3431--3440},
  year={2015}
}

@inproceedings{EffNetV1,
  title={Efficientnet: Rethinking model scaling for convolutional neural networks},
  author={Tan, Mingxing and Le, Quoc},
  booktitle={International conference on machine learning},
  pages={6105--6114},
  year={2019},
  organization={PMLR}
}

@inproceedings{EffNetV2,
  title={Efficientnetv2: Smaller models and faster training},
  author={Tan, Mingxing and Le, Quoc},
  booktitle={International conference on machine learning},
  pages={10096--10106},
  year={2021},
  organization={PMLR}
}

@inproceedings{squeeze,
  title={Squeeze-and-excitation networks},
  author={Hu, Jie and Shen, Li and Sun, Gang},
  booktitle={Proceedings of the IEEE conference on computer vision and pattern recognition},
  pages={7132--7141},
  year={2018}
}

@techreport{ATL-PHYS-PUB-2021-028,
      collaboration = "ATLAS",
      title         = "{Identification of hadronically-decaying top quarks using
                       UFO jets with ATLAS in Run 2}",
      institution   = "CERN",
      reportNumber  = "ATL-PHYS-PUB-2021-028",
      address       = "Geneva",
      year          = "2021",
      url           = {https://cds.cern.ch/record/2776782},
      note          = {All figures including auxiliary figures are available at \url{https://atlas.web.cern.ch/Atlas/GROUPS/PHYSICS/PUBNOTES/ATL-PHYS-PUB-2021-028}},
}

@article{Larkoski:2014gra,
    author = "Larkoski, Andrew J. and Moult, Ian and Neill, Duff",
    title = "{Power Counting to Better Jet Observables}",
    eprint = "1409.6298",
    archivePrefix = "arXiv",
    primaryClass = "hep-ph",
    reportNumber = "MIT--CTP-4588",
    doi = "10.1007/JHEP12(2014)009",
    journal = "JHEP",
    volume = "12",
    pages = "009",
    year = "2014"
}

@article{Larkoski:2015kga,
    author = "Larkoski, Andrew J. and Moult, Ian and Neill, Duff",
    title = "{Analytic Boosted Boson Discrimination}",
    eprint = "1507.03018",
    archivePrefix = "arXiv",
    primaryClass = "hep-ph",
    reportNumber = "MIT-CTP-4681",
    doi = "10.1007/JHEP05(2016)117",
    journal = "JHEP",
    volume = "05",
    pages = "117",
    year = "2016"
}

@article{Larkoski:2014zma,
    author = "Larkoski, Andrew J. and Moult, Ian and Neill, Duff",
    title = "{Building a Better Boosted Top Tagger}",
    eprint = "1411.0665",
    archivePrefix = "arXiv",
    primaryClass = "hep-ph",
    reportNumber = "MIT-CTP-4595, MIT--CTP-4595",
    doi = "10.1103/PhysRevD.91.034035",
    journal = "Phys. Rev. D",
    volume = "91",
    number = "3",
    pages = "034035",
    year = "2015"
}

\end{document}